\documentclass[leqno,openbib]{article}
\usepackage{indentfirst}
\usepackage{amsfonts}
\usepackage{ntheorem}
\usepackage{amsmath}
\usepackage[doublespacing]{setspace}
\usepackage{sectsty}
\usepackage[norule,bottom]{footmisc}
\usepackage[justification=centering,textfont={sc},labelfont={rm}]{caption}
\usepackage{varioref}
\usepackage{natbib}
\usepackage{graphicx}
\usepackage{hyperref}
\usepackage{makecell}
\usepackage{amsmath}
\usepackage{geometry}
\usepackage{changepage}

\theoremindent\parindent
\makeatletter    
\renewtheoremstyle{plain}{\item[\hskip\labelsep\hskip-\parindent \theorem@headerfont ##1\ ##2\theorem@separator]}{\item[\hskip\labelsep\hskip-\parindent \theorem@headerfont ##1\ ##2\ (##3)\theorem@separator]}
\makeatother
\theoremheaderfont{\scshape}
\theorembodyfont{\rmfamily}
\theoremseparator{.}

\sectionfont{\normalfont\scshape\centering}
\renewcommand{\thesection}{\Roman{section}.}
\renewcommand{\thesubsection}{\thesection\Alph{subsection}.}
\subsectionfont{\itshape}
\makeatletter
\def\@biblabel#1{\hspace*{-\labelsep}}

\makeatother
\makeatletter
\renewcommand\@makefnmark{\mbox{\textsuperscript{\normalfont\@thefnmark}}}
\renewcommand\@makefntext[1]{\indent\makebox[2.5em][r]{\@thefnmark.\,}#1}
\if@titlepage
  \renewenvironment{abstract}{      \titlepage
      \null\vfil
      \@beginparpenalty\@lowpenalty
      \begin{center}        \@endparpenalty\@M
      \end{center}}     {\par\vfil\null\endtitlepage}
\else
  \renewenvironment{abstract}{      \if@twocolumn
      \else
        \small
        \begin{center}        \end{center}      \fi}
      {\if@twocolumn\else\endquotation\fi}
\fi
\renewcommand\thetable{\@Roman\c@table}
\renewcommand\tablename{TABLE}
\renewcommand\thefigure{\@Roman\c@figure}

\makeatother
\labelformat{equation}{(#1)}
\begin{document}

\title{Improving Macroeconomic Model Validity and Forecasting Performance with Pooled Country Data using Structural, Reduced Form, and Neural Network Models}
\author{%
\textsc{Cameron Fen and Samir Undavia\thanks{Corresponding author Cameron Fen is PhD student at the University of Michigan, Ann Arbor, MI, 48104 (E-mail: camfen@umich.edu. Website: cameronfen.github.io.). Samir Undavia is an ML and NLP engineer (E-mail: samir@undavia.com). The authors thank Xavier Martin Bautista, Thorsten Drautzburg, Florian Gunsilius, Jeffrey Lin, Daniil Manaenkov, Matthew Shapiro, Eric Jang, and the member participants of the European Winter Meeting of the Econometric society, the ASSA conference, the Midwestern Economic Association conference, EconWorld conference, and the EcoMod conference for helpful discussion and/or comments. We would also like to thank Nimarta Kumari for research assistance assembling the DSGE panel data set and AI Capital Management for computational resources. All errors are our own.}}}
\maketitle

%\begin{center} \Large Please do not cite or circulate without permission\end{center}
\vspace*{-2cm}
\begin{abstract}
\begin{singlespace}

%  This abstract is 150 words:
We show that pooling countries across a panel dimension to macroeconomic data can improve by a statistically significant margin the generalization ability of structural, reduced form, and machine learning (ML) methods to produce state-of-the-art results. Using GDP forecasts evaluated on an out-of-sample test set, this procedure reduces root mean squared error by 12\% across horizons and models for certain reduced-form models and by 24\% across horizons for dynamic structural general equilibrium models. Removing US data from the training set and forecasting out-of-sample country-wise, we show that reduced-form and structural models are more policy-invariant when trained on pooled data, and outperform a baseline that uses US data only. Given the comparative advantage of ML models in a data-rich regime, we demonstrate that our recurrent neural network model and automated ML approach outperform all tested baseline economic models. Robustness checks indicate that our outperformance is reproducible, numerically stable, and generalizable across models.
\newline
% This abstract is 97 words:
% We show that pooling countries’ macroeconomic data across a panel dimension produces a statistically significant improvement in the generalizability of structural, reduced form, and machine learning (ML) methods, producing state-of-the-art results. Using GDP forecasts evaluated on an out-of-sample test set, this procedure reduces RMSE anywhere from 12\% to 24\% depending on the type of model. Forecasting using non-US-pooled-data, we show that reduced-form and structural models are more policy-invariant and outperform a US-data-only baseline. Our deep learning approaches outperform all tested baseline economic models. Robustness checks indicate that our outperformance is reproducible, numerically stable, and generalizable across models.
\end{singlespace}
\end{abstract}

%TCIMACRO{%
%\TeXButton{Titlepage formatting}{\thispagestyle{empty}\setcounter{page}{0}}}%
%BeginExpansion
\thispagestyle{empty}\setcounter{page}{0}%
%EndExpansion
\newpage 

\section{Introduction}
A central problem in both reduced-form and structural macroeconomics is the dearth of underlying data. For example, GDP is a quarterly dataset that only extends back to around the late 1940s, which results in around 300 timesteps. Thus generalization and external validity of these models are a pertinent problem. In forecasting, this approach is partially addressed by using simple linear models. In structural macroeconomics the use of micro-founded parameters and Bayesian estimation attempts to improve generalization to a limited effect. More flexible and nonparametric models would likely produce more accurate forecasts, but with limited data, this avenue is not available. However, pooling data across many different countries allows economists to forecast and even estimate larger structural models which have both better external validity and forecasting, without having to compromise on internal validity or model design.  

We show that the effectiveness of pooling US or other single country data with other countries in conjunction with large DSGE models and machine learning leads to improvements in external validity and economic forecasting. This pooling approach adds more data, rather than more covariates and leads to consistent and significant improvements in external validity as measured by timestep out-of-sample forecasts and other metrics. For example, our data goes from 211 timesteps of US data for our AutoML model, to 2581 timesteps over all countries in our pooled data. This not only leads to significant improvements in forecasting for standard models, but also allows more flexible models to be used without overfitting. Pooling a panel of countries also leads to parameters that are a better fit to the underlying data generating process -- almost for free -- without any change to the underlying equations governing the models and only changing the interpretation from a single country parameter to an average world parameter. Even in this case we show that the stability of parameters over space -- across countries -- may be better than the alternative--going back further in time for more single country data. 

A central theme throughout this paper is that more flexible models benefit more from pooling. We start with the linear models as a good baseline. Even in this case, pooling improves performance. Estimating traditional reduced-form models -- AR(2) \citep{doi:10.1098/rspa.1931.0069}, VAR(1), VAR(4) models \citep{10.2307/1912017} -- we show that we can reduce RMSE by an average of 12\% across horizons and models. Outside of pure forecasting, analysis of our pooling procedure across models suggests improvements in external validity in other ways -- making models more policy/regime invariant. To show this, we estimate both linear and nonlinear models on data from all countries except the target country being forecasted. Thus, our test data is not only time step out-of-sample, which we implement in all our forecast evaluations, but also country out-of-sample, which we add to illustrate generalizability. Across most models and forecasting horizons, our out-of-sample forecasts outperforms the typical procedure of estimation models on only the data of the country of interest. This time and country out-of-sample analysis leads to about 80\% of the improvement gained from moving all the way to forecasting with the full data set. We believe this provides evidence that this data augmentation scheme can help make reduced-form as well as more nonlinear models more policy-invariant. 

Moving to a more flexible model, we proceed to apply our panel data augmentation scheme to the Smets-Wouters structural 45+ parameter DSGE models \citep{smets2007shocks}. Our augmentation statistically improves the generalization of this model by an average of 24\% across horizons. We again test our model in  a country out-of-sample manner, and show improvements while estimating over a single country baselines. This suggests that even DSGE models are not immune to the Lucas critique and the use of country panel data can improve policy/country invariance. Given the consistent improvements across all horizons and reduced-form models, we are confident this approach will generalize to the estimation of other structural models. We advocate applying this approach to calibration, generalized method of moments \citep{hansen1982generalized}, and Bayesian estimation \citep{fernandez2007estimating}, where the targets are moments from a cross-section of countries instead of just one region like the United States.

Recognizing that this augmentation increases the effective number of observations by a factor of 10, we also demonstrate that pooling can overcome overfitting in flexible machine learning models that can now outperform traditional forecasting models in this high data regime. This is in line with the trend of larger models having a comparative advantage in terms of forecasting improvement given more data. We use two different algorithms. The first, AutoML, runs a horse race with hundreds of machine learning models to determine the best performing model on the validation set, which is ultimately evaluated on the test set. We view AutoML as a proxy for great machine learning performance, but also expect individual model tuning can improve performance even further. As different models perform better under the low-data (US) regime and the high-data (pooled) regime under AutoML, we also test an RNN to show the improvement of a single model under both data regimes. The model improvement indicates that while these approaches are competitive in the low data regime, machine learning methods consistently outperforms baseline economic models -- VAR(1), VAR(4), AR(2), Smets-Wouters, and factor models -- in the high data regime. Furthermore, while some of the baseline models use a cross-section of country data, we only use three covariates -- GDP, consumption, and unemployment lags. In contrast, the DSGE model uses 11 different covariates, the factor model uses 248, and the Survey of Professional Forecasters (SPF) \citep{SPF} uses just as many covariates along with real time data, suggesting our machine learning models still have room for improvement. Over most horizons, our model approaches SPF median forecast performance, albeit evaluated on 2020 vintage data (see Appendix \ref{spf_appendix}), resulting in outperformance over SPF benchmark at 5 quarters ahead. Moreover, the outperformance of our model over the SPF benchmark is noteworthy as the SPF is an ensemble of both models and human decision makers, and many recessions like the recent COVID-19 recession are not generally predicted by models.

The paper proceeds as follows: Section \ref{lit_review_section} reviews the literature on forecasting and recurrent neural networks and describes how our paper merges these two fields; Section \ref{section_automl} discusses feed-forward neural networks, linear state space models, and gated recurrent units \citep{DBLP:journals/corr/ChoMBB14}; Section \ref{section_data_and_method} describes the data; Section \ref{section_our_model} briefly mentions our model architecture; Section \ref{section_economic_models} discusses the benchmark economic models and the SPF that we compare our model to; Section \ref{section_results} and Appendix \ref{robustness_appendix} provide the main results and robustness checks; and Section \ref{section_conclusion} concludes the paper. 

\section{Literature Review}
\label{lit_review_section}
This paper connects multiple strands of literature: machine learning, time-series econometrics, and panel macroeconomic analysis.  

As our pooling technique leads to larger datasets, this creates an opportunity to either increase the parameter count in models or proceed to more powerful tools. In the area of machine learning, when combined with additional data, even models with billions of parameters still exhibit continued log linear improvements in accuracy \citep{DBLP:journals/corr/abs-2001-08361}. This opens up an avenue to explore whether the outperformance of linear models is due more to 1) the lack of data or 2) their attractive properties in fitting the underlying data generating process. The results of pooling across countries suggests that the advantage of linear models is due to the former.

We applied a recent machine learning technique to our economic data, AutoML, a technique introduced and improved on as machine learning models became more complicated with more layers and hyperparameters to tune \citep{thornton2013auto}. Unlike the case with deep learning, many innovations came from the software industry to automate estimation techniques. However, there is a vibrant academic literature following its introduction \citep{automlws}. The basic premise is to automate the model training and discovery portion of machine learning. H2O \citep{ledell2020h2o}, the AutoML platform we use, takes data and automatically runs a horse race of machine learning models on the validation set then returns the best performing model. Hence, we view the output model as a proxy for a well trained and effective predictive model by a data scientist, even if some additional fine tuning can improve performance. While there is room for human improvements over a automated machine learning process, removing the human from the process entirely in our AutoML algorithm shields us from most p-hacking critiques. 

The second machine learning procedure we used was the estimation of a RNN, which is a state space model much like the linear state space models often used in economics. Innovations in deep learning have improved the predictive power of these models over what economists are used to for their linear counterparts. RNNs have been around in many forms, but were mainly popularized in the 1980s \citep{rumelhart1986learning}. The popularity and performance of RNNs grew with the introduction of long short-term memory (LSTM) networks by \citet{Hochreiter:1997:LSM:1246443.1246450}. The model uses gates to prevent unbounded or vanishing gradients giving this model the ability to have states that can ``remember'' many timesteps into the past. In addition to its pervasive use in language modeling, LSTMs are used in fields as disparate as robot control \citep{Robot_Control_LSTM}, protein homology detection \citep{Protein_Detection_LSTM}, and rainfall detection \citep{Rainfall_LSTM}. We use a modification of long short-term memory networks called a gated recurrent unit \citep{DBLP:journals/corr/ChoMBB14}. RNNs and other deep learning architectures like convolutional neural networks have been used to forecast unemployment \citep{smalter2017macroeconomic}. Within economics, gated recurrent units (GRUs) have been applied in stock market prediction \citep{minh2018deep} and power grid load forecasting \citep{li2020short}. 

Moving from machine learning models to economics models, autoregressive models have been the workhorse forecasting models since the late 1930s \citet{diebold1998elements}, \citet{doi:10.1098/rspa.1931.0069}). Even the machine learning models maintain an autoregressive structure in its inputs. Despite its simplicity and age, the model is still used among practitioners and as a baseline in many papers \citep{watson2001time}. One advancement in forecasting stems from the greater adoption of structural or pseudo-structural time series models like the Smets-Wouters DSGE models \citep{smets2007shocks}. While DSGE forecasting is widely used in the literature, they are competitive with -- but often times no better than -- a simple AR(2), with more bespoke DSGE models performing poorer \citep{edge2010comparison}. However, the use of DSGE models for counterfactual analysis is an important and unique benefit of these models. The final economic baseline is the factor model \citep{stock2002forecasting}, which attempts to use a large cross-section of data resulting in a more comprehensive picture of the economy to perform forecasting.

Details on all these models and our implementation can be found in Appendix \ref{econ_models_appendix}. In addition, our paper uses tools from forecast evaluation \citep{west1996asymptotic}, \citep{pesaran1992simple}, and \citep{diebold2002comparing}, as well as model averaging \citep{koop2012bayesian}, \citep{timmermann2006forecast}, and \citep{wright2008bayesian}. 

Moving to structural economics, there is a scant but robust literature on panel data and dynamic general equilibrium models \citep{breitung2015analysis}. Most of the literature focuses on the use of panel data to better identify the effects of interest that vary across country. Much of it is theoretical and adopts microeconometric panel techniques to macroeconomic data \citep{banerjee2004some}. This literature also studies the use of cross-sectional data to improve counterfactual analysis in general equilibrium models \citep{miyamoto2018government}, \citep{crandall2007effects} to have a more microeconomic forecasting focus \citep{baltagi2008forecasting}. There is also literature looking at specific panel models applied to macroeconomics like dynamic panel models \citep{doran2006gmm}, \citep{bai2010instrumental}, and \citep{diebold2006macroeconomy}. 

At the same time, the approach of pooling countries has faced some resistance for theoretical reasons. \citet{pesaran1995estimating} argue that structural parameters lose meaning as they turn into a mean value across countries rather than an estimate for the true value in one country. However, our results suggest that even if using a spatial dimension across countries, the econometrician still needs a minimum amount of data for good parameter identification. If one pools across a large spatial cross-section one can use data that is more recent. As we show empirically, more recent data but spread across different countries has the same -- if not -- more predictive power than data that is constrained to a single country but extends further into the past. This finding suggests that even though it might be neater to use single country data extending further back in time, countries are somewhat artificial boundaries. The stability and predictive power of parameters are at least as strong across space as across time. 

\section{Machine Learning Models}
\label{section_automl}

\subsection{Automated Machine Learning}
AutoML software is designed to provide end-to-end solutions for machine learning problems by efficiently training and evaluating a number of different models and ultimately returning the best model. In order to provide a proxy for the performance of a good ``nonparametric'' machine learning model, we tested the open-source automated machine learning (AutoML) software H2O\footnote{https://www.h2o.ai/}. We created a pipeline for each prediction horizon, trained the model using our international cross-sectional data, evaluated on US validation data, and lastly predicted using our US data test set. In contrast with our own custom model, setting up H2O and training on our dataset was almost entirely automated. 

The benefit of automation is that while humans can improve performance, there was little we could do either via tinkering with architecture or devoting more computational resources to influence the performance of the procedure. From predicting one quarter ahead to five quarters ahead, the AutoML software picked a different model for each horizon, respectively: XGBoost, gradient boosting machine, gradient boosting machine, distributed random forest, and deep learning. We noticed that the software generally picked deep learning models for the quarters that were further away (four and five quarters ahead) compared to predicting gradient boosted techniques for closer quarters (one and two quarters ahead). This is not surprising, as decision-tree-based techniques have relatively few degrees of freedom and good inductive biases for many problems but deep learning techniques ultimately are more flexible and scale better on larger datasets because of larger parameter counts. Ultimately, AutoML had very strong results and can be applied to other prediction problems in economics.

Additionally, because the AutoML selects a different model for a given horizon and data set size, we also estimated an RNN on the both the reduced USA dataset and the pooled world data set. This allows us to show the effect of the increase in data size holding the model architecture fixed. The RNN also has the advantage of not being a model considered by AutoML, which gives broader coverage of the universe of machine learning models that are being considered in our paper.

% We spend a good deal of time designing an architecture that would work effectively for this task. The details of the model are contained in Appendix N. Aside from standard improvements like Adam \citep{Adam}, batch normalization \citep{ioffe2015batch}, and skip connections \citep{he2015deep}, we introduce an innovative structure in the last layer using the input to skip the entire network and concatenating the output of the network with the original inputs and using a final linear layer to perform the regression prediction. This innovation allows the neural network to nest AR and VAR models. Other than that training is standard.

\subsection{Our Neural Network Model Architecture}
\label{section_our_model}
\begin{figure}[htbp]
    \centering
    \includegraphics[width=10cm]{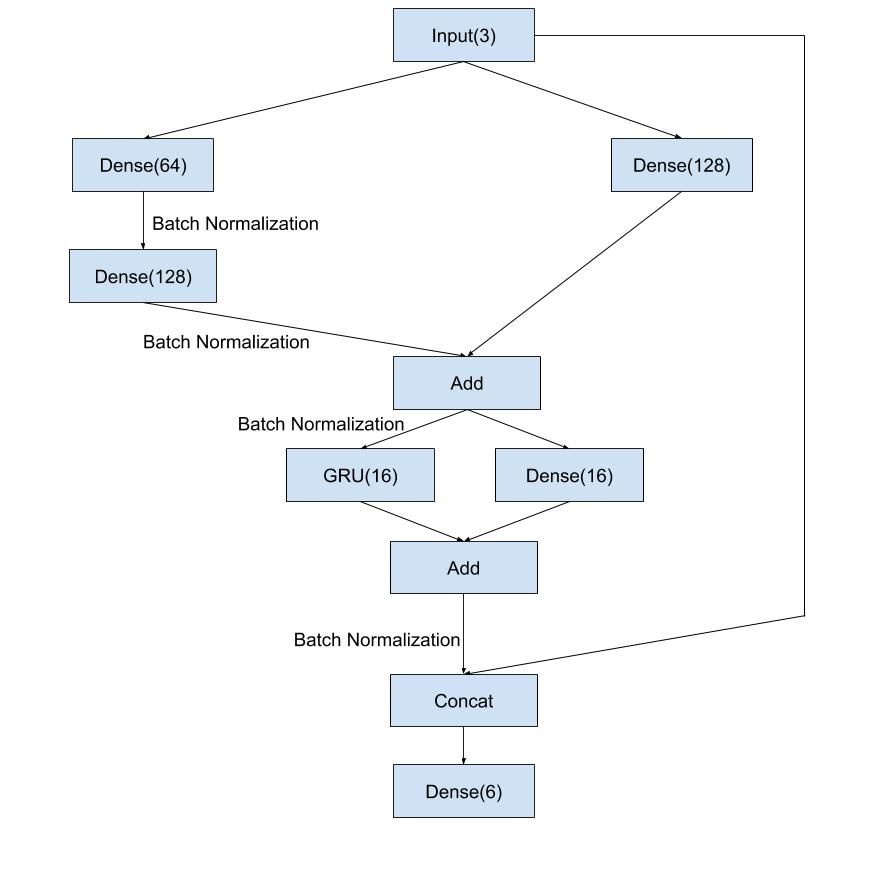}
    \caption{Model Architecture}
    \label{fig:Model Architecture}
\end{figure}

A RNN model we use to supplement AutoML is the gated recurrent unit (GRU) model, described in Appendix \ref{gru_appendix}. We add additional feedforward layers as well as other architecture choices as indicated in Figure \ref{fig:Model Architecture}. The model architecture involves preprocessing the data using two feed-forward dense layers with 64 and 128 hidden units respectively and rectified linear unit \citep{DBLP:journals/corr/abs-1803-08375} activation (see Appendix \ref{relu_appendix}.), then running a gated recurrent unit with sixteen states on this preprocessed data.\footnote{While we use the word \textit{preprocessed}, the approach is trained entirely end-to-end and is not a two step process as the word preprocess might imply. The neural network projects the input data -- consumption growth, GDP growth and the unemployment rate -- into a high dimensional state that the gated recurrent unit finds easier to work with much like pre-processing would. The end-to-end procedure learns the pre-processing and the gated recurrent unit analysis at the same time} Finally the output of the GRU is concatenated with the original input (lagged GDP, consumption and unemployment) and fed into a linear layer which forecasts an output. 

Our model contains parallel dense layers between each operation; the layers were originally skip connections \citep{he2015deep}, but we modified them to allow for learning of linear parameters. The final skip connection concatenates the input with the output of the network so that the neural network would nest an VAR(1) model. These design choices all improved forecasting performance. Between all of our non-skip connection layers, we also use batch normalization \citep{ioffe2015batch}. More details on batch normalization can also be found in Appendix \ref{batch_norm_appendix}. Ultimately, our model comprises about 17,000 parameters which explains the comparative outperformance on a data rich regime.

\section{Economic Models}
\label{section_economic_models} 
%TODO Rewrite this paragraph with more formal language
We tested the predictive power of a series of machine learning and traditional macroeconomic models estimated on our panel of countries using our novel data pooling method. We found that the more complex the model, the more our data augmentation helped. The machine learning models tended to be more flexible, but even among economic models the trend still held. Additionally, we provided comparisons to the Survey of Professional Forecasters \citep{SPF} median GDP forecast, which is seen as a proxy for state-of-the-art performance. A discussion of the Survey of Professional Forecasters and our attempt to evaluate their forecasts is contained in Appendix \ref{spf_appendix}. The baseline economic models we used are the AR(2) autoregressive model, the Smets-Wouters 2007 DSGE model \citep{smets2007shocks}, and a factor model \citep{stock2002forecasting}, \citep{stock2002macroeconomic} and a VAR(4)/VAR(1) \citep{10.2307/1912017}. A more detailed explanation of these models is contained in Appendix \ref{econ_models_appendix}. For the linear models, getting cross country data is straightforward, thus we compare those models estimated only on US data as well as on our data set of 50 countries. For the Smets-Wouters DSGE, we also assembled a panel of 27 rich and developing countries to estimate the structural model on.

As is standard with economic forecasting, the baseline models were trained in a pseudo-out-of-sample fashion where the training set/validation set expands as the forecast date becomes more recent. However, with our neural network and AutoML, we keep the training set and validation set fixed due to computational costs and programming constraints. We expect that our model will improve if we use a pseudo-out-of-sample approach.

\section{Data and Method}
\label{section_data_and_method}
When initially training our complex neural network models, we found that United States macroeconomic data was not sufficient so, in order to train the model, we use data from 49 other developed and developing countries as listed in Appendix \ref{countries_appendix}. We source cross country data from Trading Economics via the Quandl platform API\footnote{https://www.quandl.com/tools/api} as well as GDP data from the World Bank.\footnote{World Development Indicators, The World Bank} We used GDP, consumption, and the unemployment rate as inputs to the model. GDP and consumption were all expressed in growth rates. Unemployment was expressed as a rate as well. As mentioned earlier, we also assembled 11 different covariates across 27 countries for a panel of data used to estimate the Smets-Wouters DSGE. Data came from the Federal Reserve Economic Data (FRED), World Bank, Eurostat, the Organization for Economic Cooperation (OECD) and the International Monetary Fund (IMF).

We split our data into training, validation, and test sets. We forecast GDP, and evaluated with RMSE. The test set was from 2008-Q4 to 2020-Q1, either testing only US data or world data depending on the particular problem. The validation consisted of data from 2003-Q4 to 2008-Q3, which was only used for the RNN. This data was in the training set for all other models. AutoML, which was not a sequential model, used k-fold cross validation on the entire training set, comprised of the remainder of data not in test or validation sets. We chose these periods so that both the test set and the validation set would have periods of both expansion and recession, based on the US business cycle. Including the 2001 recession in the validation set would leave the model without enough training data, so we split the data of the Great Recession over the test and validation set. The quarter with the fall of Lehman Brothers and the largest dip in GDP was the first quarter in our test set, 2008-Q4. Two quarters with negative growth preceding this were in the validation set. We estimated all models from a horizon of one quarter ahead to five quarters ahead. The metric of choice for forecast evaluation was RMSE. 

\section{Results} \label{section_results}

%put somewhere

%end inserted paragraph
Our first set of results shows the benefit of pooled data on reduced-form models VARs and ARs, showing significantly improved GDP forecasting accuracy. Despite the relative parsimony of these models adding pooled data improves RMSE performance almost uniformly by an average of 12\%. Our second set of results shows that the panel data augmentation improved the chances of building externally and potentially, internally valid structural models, using the Smets-Wouters DSGE model as the main benchmark. These models benefited more due to the pooled data as they had a higher parameter count, improving RMSE by 24\%. Our third set of results took all the models and demonstrates the forecasting power of ``nonparametric" machine learning models over all the previously mentioned traditional economic models in this relatively data rich regime. As these models were the largest and most data hungry, the use of pooled data improved performance from slightly above average forecasters to providing state-of-the-art predictions across the board. The RMSE of the RNN-based model improved by 23\%, which was smaller than the structural models. However, the RMSE starting point was much better than that of the structural models and proportional performance was more difficult the better the original baseline. Ultimately, we present an interesting finding in the improvement in performance as model capacity increased, moving from the US data set to the pooled world data set. This suggests that more complex models benefit from increasingly larger datasets and that pooling can address overfitting.

\subsection{Reduced-Form Models}

\begin{figure}[htbp]
    \centering
    \includegraphics[width=10cm]{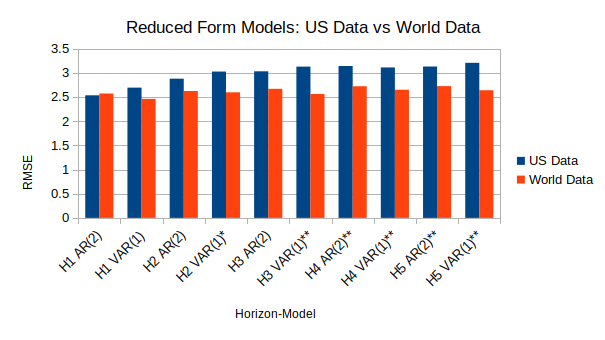}
    \caption{Evaluation of RMSE for Linear Models Using Both US Only Data and Pooled World Panel Data}
    \label{fig:RF RMSE US vs World}
\end{figure}

Figure \ref{fig:RF RMSE US vs World} shows the forecasting performance of either a VAR(1) or AR(2) model at a particular forecast horizon. The first bar in each pair represents the forecast RMSE on the test set (2008Q4-2020Q1) with the model estimated only on US data. The second bar in each pair represents forecasting performance using the model estimated on the entire panel of countries. The stars next to the model's name indicates the statistical significance of world data outperformance over the US data using a Diebold-Mariano test \citep{diebold2002comparing} at the 1\% (***), 5\% (**) or 10\% (*) level. This format will be followed for the rest of the RMSE performance graphs, unless otherwise noted. 

Pooling improved the performance of the models in statistically significant manner, especially at longer horizons. Except for a slight underperformance at the one quarter ahead for the AR(2), all other horizon-models show outperformance using the country panel data augmentation. The outperformance of the pooled data averages roughly 12\% of US RMSE over all horizon-model pairs. We show a significant improvement with the pooled data, however, since the set has limited model complexity, the improvement is not as large as that of more complex structural or machine learning models.

\begin{figure}[htbp]
    \centering
    \includegraphics[width=10cm]{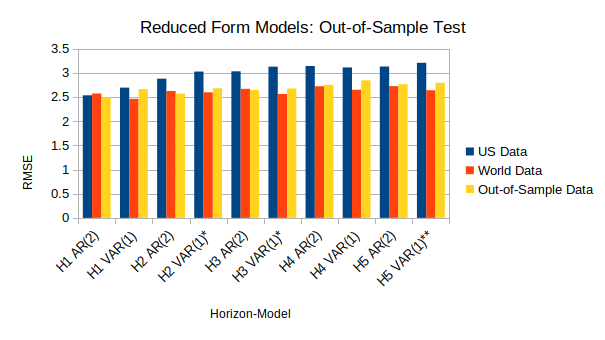}
    \caption{Reduced-Form Time Out-of-Sample as Well as Country Out-of-Sample Forecasts}
    \label{fig:RF RMSE OOS}
\end{figure}

Figure \ref{fig:RF RMSE OOS} shows similar data to Figure \ref{fig:RF RMSE US vs World}, except the third bar for each horizon-model triplet shows forecasting performance on the same US test data, but for a model that is both time step and country out-of-sample. For example, since we forecasted US GDP, we used every country but the US to forecast US GDP. This test enables us to show that using panel data can lead to models that are policy/country invariant and can generalize even to country data that the model lacked access to. In all cases, the RMSE of the out-of-sample forecasts were statistically indistinguishable from the RMSE of the model estimated on the full panel of countries but significant over the US baseline with stars indicating significance of the out-of-sample forecast. Again the models were mainly significant at longer horizons, but any significance is nevertheless impressive since the outperforming model uses no US data to forecast US GDP, for example. Excluding the H1 AR(2) pair and using only the out-of-sample data led to capturing 79\%, on average, of the outperformance of the world panel over US only data forecasts. Ultimately, there was an improvement in performance due to the use of the pooled data over single country training data.

\newcommand*{\MyIndent}{\hspace*{0.5cm}}
\begin{table}[htbp] \centering
\caption{Average Forecasting Performance Evaluated over the World}
%uncertainty of multiple runs factored in the significant figures
\label{World Forecasting Performance}
\begin{tabular}{cccccc}
\hline \hline
% \multicolumn{6}{ c }{Root Mean Squared Error}\\
\multicolumn{1}{r}{Time (Q's Ahead)}   & 1Q           & 2Q           & 3Q           & 4Q           & 5Q           \\ \hline
\multicolumn{1}{l}{AR(2)} \\
\multicolumn{1}{l}{\MyIndent Local Data}         & 5.22          & 5.33          & 5.51         & 5.56          & 5.60          \\
\multicolumn{1}{l}{\MyIndent World Data}         & 4.88          & 4.98          & 5.10         & 5.19          & 5.19          \\ 
\multicolumn{1}{l}{\MyIndent Out-of-Sample Data}         & 5.07          & 5.13          & 5.36         & 5.36          & 5.38          \\ \hline
\multicolumn{1}{l}{VAR(1)} \\
\multicolumn{1}{l}{\MyIndent Local Data}         & 5.10          & 5.17          & 5.19         & 5.29          & 5.26          \\
\multicolumn{1}{l}{\MyIndent World Data}         & 4.80          & 4.94          & 5.04         & 5.11          & 5.12          \\ \multicolumn{1}{l}{\MyIndent Out-of-Sample Data}         & 4.92          & 5.00          & 5.07         & 5.20          & 5.25          \\ \hline
\multicolumn{1}{l}{VAR(4)} \\
\multicolumn{1}{l}{\MyIndent Local Data}         & 7.90          & 7.05          & 7.74         & 7.87          & 9.27          \\
\multicolumn{1}{l}{\MyIndent World Data}         & 4.72          & 4.90          & 5.03         & 5.10          & 5.11          \\ \multicolumn{1}{l}{\MyIndent Out-of-Sample Data}         & 4.70          & 4.87          & 5.02         & 5.17          & 5.26          \\\hline\hline\\
\end{tabular}
\end{table}

We performed the same tests over our entire cross-section of countries, in Table \ref{World Forecasting Performance}. For example, for each local forecast, we used only French data to forecast French GDP. Likewise, for world forecasts, we used the entire data set to estimate a single model, then for each country we used the same model to jointly forecast the GDP of every country and average RMSE values across countries. For out-of-sample data, a different model is trained to forecast each country, but each model is estimated using every country but the country of forecasting interest. Aside from providing an additional robustness test regarding the outperformance of world data and even out-of-sample data, this table shows additional policy-invariance both from the out-of-sample tests as well as the world data tests, where the same model can jointly forecast all countries better than custom tailored models for each country. Additionally, this model seems to indicate a diminishing return in terms of RMSE for linear models when using pooling data and it seems like all linear models converge to a similar RMSE, although the larger VAR(4) models seem to have the best performance and suggests improvement when moving to more complex structural and machine learning models.

\subsection{Structural Models}
Considering the success of data pooling for reduced-form models, we also tested the procedure on structural models and achieved even greater success. Since the DSGE model requires 11 different variables, we assembled our own data from the World Bank, OECD, Eurostat, and FRED. This panel has only a 27 country cross-section which is smaller than the 50-country panel used in our reduced-form models. The performance of our structural models demonstrates that this pooling of data likely leads to performance gains across models, including DSGE models that should generalize to out-of-sample data because of their resilience to the Lucas critique. Combined with the results using machine learning models, our results make the case that model outperformance due to pooling helps across many, if not most, types of models. The three results we show entails: 
\begin{adjustwidth}{20pt}{20pt}
1) Comparing the forecasting performance of a DSGE model trained only on US data to a DSGE model estimated on the entire panel of countries in the same manner as Figure \ref{fig:RF RMSE US vs World},  \newline
2) Comparing the performance of a model that is not only out-of-sample from a timestep perspective as is usual in forecasting, but also does not contain the forecasting country data in question when estimating, in the same manner as Figure \ref{fig:RF RMSE OOS}; and \newline
3) Illustrating an empirical evidence testing parameter stability across time versus across space and showing that parameters are likely as predictive -- if not more predictive -- when working across space as across time, which suggests that pooling may not have an interpretive disadvantage over estimating data much further back in time from only one country. 
\end{adjustwidth}
The first result is illustrated in this graph below,

\begin{figure}[htbp]
    \centering
    \includegraphics[width=10cm]{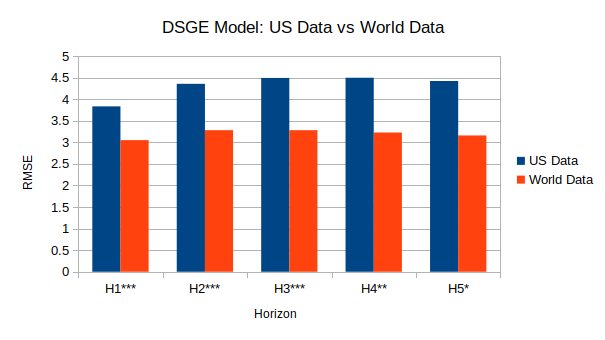}
    \caption{Evaluation of RMSE for DSGE Models Using Both US Only Data and Cross-Country Panel Data}
    \label{fig:DSGE RMSE US vs World}
\end{figure}

Unlike the reduced-form models, the most significant results for structural models, with p-values less than 1\%, were at shorter horizons -- below three quarters ahead -- although all horizons were significant at the 10\% level. The parameters that seemed to change the most -- moving from US only data to world data -- were the shocks and the moving averages of the shock variable, monetary policy Taylor-rule variables, and variables governing wages and inflation. While the increasing variance of the shocks did not affect expected forecasts due to certainty equivalence, the model is both less confident and closer to correct when using pooled world data. Perhaps it is unsurprising that variables focusing on monetary policy and inflation are different when estimated on world data. Inflation, especially among rich developing countries, along with the monetary response to them, was a more pernicious problem outside the US than within \citep{azam2020threshold}. For more information on the changes in structural variables when moving from US data to pooled world data see Appendix \ref{param_comp}. 

Despite the increased uncertainty of the model as illustrated by the increase in the standard deviation of the shocks, the parameters were more reliable when estimated under world data. The improvement in RMSE averages over 25\% over all horizons, which was more than double the percentage improvement for reduced-form models. Part of the outperformance was due to weaker performance of the models estimated on US data. This suggests that the Smets-Wouters model is no better at generalizing across policy regimes or countries than reduced-form models and benefits more because of its higher parameter count.

\begin{figure}[htbp]
    \centering
    \includegraphics[width=10cm]{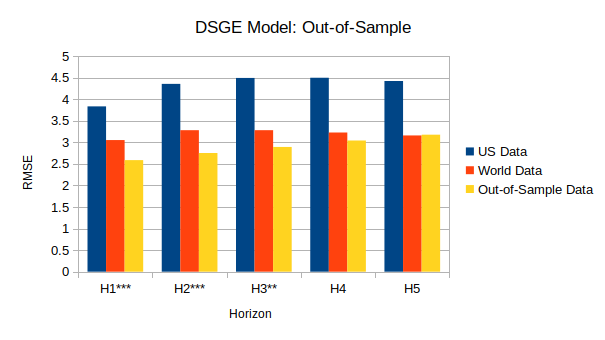}
    \caption{DSGE Time Out-of-Sample as Well as Country Out-of-Sample Forecasts}
    \label{fig:DSGE RMSE OOS}
\end{figure}

In Figure \ref{fig:DSGE RMSE OOS}, we also provide a structural chart that parallels the out-of-sample chart in Figure \ref{fig:RF RMSE OOS} in the reduced-form section. As a reminder, the out-of-sample DSGE model was estimated on the panel of 26 countries (removing the US) so that the GDP forecast was country out-of-sample as well as time step out-of-sample. The out-of-sample performance was 9\% better, on average, than even the performance of a DSGE estimated on the entire world data. However, the Diebold-Mariano tests are less significant with only the first two horizons having p-values with less than 1\% and no significance in horizon four and five. This suggests that the out-of-sample outperformance may in part be due to chance, and we studied this in further detail below. Since the United States was the only country which has data that extends much before 2000 across all our needed variables, we hypothesized that DSGE model parameters were more stable across countries than across time. Removing the US made the data closer in time to the test set. This addresses an internal validity criticism of our panel approach arguing that when pooled structural parameters have different values across countries, estimating a single model on all countries strips the parameters of economic meaning \citep{pesaran1995estimating}. For example the parameter no longer represents the depreciation rate of the United States, but an average depreciation rate across 30 countries. This result provides a suggestive counter argument to that claim, by pointing out that, considering one needs a certain number of data points to get accurate predictions to begin with, using data across a cross section of countries provides forecasts that at least as good as using data the extends further in time. 

\begin{figure}[htbp]
    \centering
    \includegraphics[width=10cm]{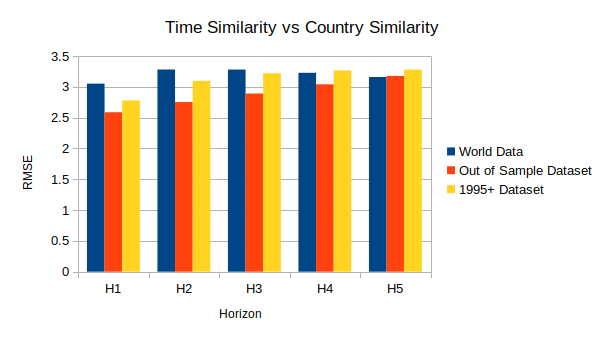}
    \caption{This Graph Compares Parameter Stability via Forecasts Going Back in Time Versus Space}
    \label{fig:DSGE Time vs Country}
\end{figure}

Probing this hypothesis led to results that were somewhat mixed. We estimated a model trained on the entire panel of countries with data from 1995-Q1 onward. This affected three countries -- the US, Japan and New Zealand. This procedure isolates more sharply  the effect of similarity across space versus across time on model generalization, rather than just removing all US data. The US lost about 140-190 data points (as the test set requires rolling forecasts), and New Zealand and Japan both lost about 15-60 timesteps. Figure \ref{fig:DSGE Time vs Country}, illustrates this experiment to compare the performance of models estimated on data since 1995 to models estimated with full country and timesteps. 

Figure \ref{fig:DSGE Time vs Country} shows GDP forecasts on the same test set: 2008Q4-2020Q1. The first bar in each triplet is the DSGE model performance estimated using world data across all periods in time. The second bar is the out-of-sample performance as shown in Figure \ref{fig:DSGE RMSE OOS}. The third bar is the performance of a DSGE model estimated on all country data, but only since 1995 to make the data more relevant across time to the test set. The 1995 onwards data performed worse than the out-of-sample test which could suggest some of the outperformance of the out-of-sample DSGE model was due to chance. However, it seems that the 1995 data performed at least as well as a model estimated on world data both pre and post-1995, despite our robust results suggesting that more data is generally better. This makes some practical sense when considering something like the advent of software, the depreciation rate in France in 2015 plausibly shares more in similarities with the depreciation rate in the US in 2020 than the depreciation rate of the US in 1960. As one needs a set number of data points to identify structural models anyway, our results provide suggestive evidence that getting cross-sectional data results in parameters that are no less stable than parameters going back in time. The results seem inconclusive but certainly don't suggest any more parameter stability across time than space, in contrast with the potential pitfall highlighted by \citet{pesaran1995estimating} and generally accepted in the literature.  

The DSGE models we estimated generally underperformed the parameters recovered in the original \citet{smets2007shocks}, because we focused on maximum likelihood estimation without priors, optimized only with gradient descent. Despite the forecasting success of \citep{smets2007shocks} and other Bayesian DSGE models \citep{herbst2015bayesian}, \citep{fernandez2016solution}, we chose to use to use a maximum likelihood approach to maintain comparability to the reduced-form and machine learning experiments as well as a large portion of the applied literature that focuses on point estimate techniques ranging from calibration, maximum likelihood, to generalized method of moments \citep{hansen1982generalized}. However, Figure \ref{fig:DSGE RMSE OOS} as well as Appendix \ref{smets_appendix} shows the performance of a model estimated via maximum likelihood outperformed along some horizons the parameters of \citet{smets2007shocks}. Given the limitations of maximum likelihood and our differing focus, we see this outperformance as an endorsement of the use of the pooling approach. We show that the use of pooled data results in the Smets-Wouters DSGE model outperforming DSGE models estimated only on US data. Furthermore, we provided suggestive evidence that is possible that models are more externally and internally valid if one uses data across countries in addition to the statistically significant improvement in generalization. Using our out-of-sample tests, we show that these models can improve the out-of-sample generalization of the Smets-Wouters DSGE model even if they are theoretically policy-invariant. The data shown in the charts are also be displayed in Appendix \ref{smets_appendix}. As a final note, while it is difficult to quantify improved performance of calibration and generalized methods of moments, based on the generalization improvements from estimation for both reduced-form and structural models, we imagine these results should generalize and macroeconomists would benefit from calibrating to moments as well as other methods that feature a large number of countries. 

\subsection{Nonparametric Machine Learning Models}
\label{nonpar_ml_models_section}
Given the improvement in forecasting performance for both the reduced-form and structural models and the improving relative performance of complex models, we decided to test the performance of nonparametric models that are even more flexible than the DSGEs and some of the larger VARs. We tested both a RNN, as well as an AutoML algorithm. While the improvement in performance was less than the DSGE improvement from pooled data, it still seems more impressive given that the flexible models had much better performance even on US data. This again illustrates the trend that increased parameter count leads to a gain in performance that came from pooling.

Two charts below illustrate performance of the RNN and AutoML models on both US and pooled world data.

\begin{figure}[htbp]
    \centering
    \includegraphics[width=10cm]{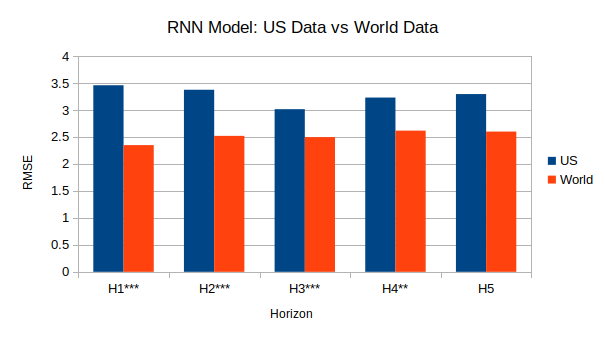}
    \caption{Evaluation of RMSE for RNN models Using Both US Only Data and Cross-Country Panel Data}
    \label{fig:RNN RMSE US vs World}
\end{figure}

We compared estimation on US data as well as pooled world data, for both models. For the RNN, Figure \ref{fig:RNN RMSE US vs World} shows the improvement in RMSE from estimating a recurrent neural network using only US data to using the entire cross-section of 50 countries. The improvement was statistically significant for all horizons except five quarters ahead. The average improvement was around 23\% over all horizons, which was similar to the improvement for the Smets-Wouters model and almost double the improvement of linear models. This is a reassuring confirmation as the RNN is a data hungry model that benefits more from data rich regimes. We also attempted to add a country identifier term to our model. For example, using GDP per capita at the time of prediction as an input to localize the pooled data to some degree. While this might be expected to reduce bias, it didn't improve out of sample performance to any degree. This potentially suggests that countries are more similar than different and the bias of pooling different countries has a limited effect, while adding in such a covariate leads to more overfitting. 

\begin{figure}[htbp]
    \centering
    \includegraphics[width=10cm]{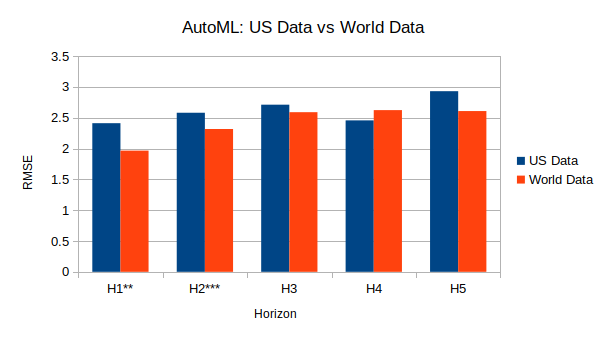}
    \caption{Evaluation of RMSE for AutoML Using Both US Only Data and Cross-Country Panel Data}
    \label{fig:AutoML RMSE US vs World}
\end{figure}

Our second chart in Figure \ref{fig:AutoML RMSE US vs World}, shows the same performance graph for AutoML. The performance gain is not as easily interpreted as AutoML benefits from the pooled data but can also pick different models that gains relatively in both data poor and data rich regimes. Because of that, the large gains of the RNN are more representative of performance gains from moving to pooled data on a fixed machine learning models. It has the least improvement in performance when using the panel of countries as training data, with average improvements in RMSE of about 7.5\%. Only the first two horizons show statistically significant forecast performance. It's worth pointing out that the improvements are still significant along some horizons, especially in light of nearly state-of-the-art performance using only US data to begin with. Even so, only in four quarters ahead does the AutoML model outperform all the baseline models trained on US data. When estimated on world data, AutoML outperforms all economic models on all horizons except three quarters ahead and its performance at one quarter ahead rivals the Survey of Profession Forecasters, despite being estimated only on lags of GDP, consumption, and unemployment and no real time data.  

\subsection{Summary} \label{results_summary_section}
The previous sections outlined performance of all reduced-form, structural, and machine learning models. This section takes all the data and provides results from a holistic perspective. We first compare forecasting models using all approaches, estimated on both pooled and US data. This table demonstrates the effectiveness of the machine learning forecasting methods in data rich regimes. AutoML estimated on world data outperforms all baseline economic models on four of the five horizons and the RNN outperforms on longer term horizons. We do not report the maximum likelihood of the Smets-Wouters model, as the original Bayesian parameterization has better performance than either of our maximum likelihood Smets-Wouters models estimated on world or US data. Introducing our other DSGE variations, would be difficult to justify and would also have no effect on the results of the horse race. Regardless, all of our models that outperform all baseline models on a horizon are bolded. No baseline model ever outperformed both our models along any horizon and the best performing model along all horizons was either an AutoML model or an RNN model, likely because the additional pooled data allowed a more powerful model to be used without overfitting. %why this is

\begin{table}[htbp] 
\label{rmse_table}
\centering
\caption{RMSE of Our RNN, AutoML, and Baseline Models}
%uncertainty of multiple runs factored in the significant figures
\scalebox{0.9}{%
\begin{tabular}{cccccc}
\hline \hline
% \multicolumn{6}{ c }{Root Mean Squared Error}\\
\multicolumn{1}{r}{Time (Q's Ahead)}   & 1Q           & 2Q           & 3Q           & 4Q           & 5Q           \\ \hline
\multicolumn{1}{l}{VAR(4)} \\
\multicolumn{1}{l}{\MyIndent US Data}         & 2.99          & 3.03          & 3.10         & 3.08          & 3.08          \\
\multicolumn{1}{l}{\MyIndent World Data}         & 2.37          & 2.52          & 2.56         & 2.63          & 2.63        \\ \hline
\multicolumn{1}{l}{AR(2)} \\
\multicolumn{1}{l}{\MyIndent US Data}         & 2.53          & 2.88          & 3.03         & 3.14          & 3.13          \\
\multicolumn{1}{l}{\MyIndent World Data}         & 2.57          & 2.62          & 2.67         & 2.72          & 2.72          \\ \hline
\multicolumn{1}{l}{Smets-Wouters DSGE Bayesian} \\
\multicolumn{1}{l}{\MyIndent US Data}  & 2.79          & 2.95          & 2.89          & 2.80          & 2.71         \\ \hline
\multicolumn{1}{l}{Factor} \\
\multicolumn{1}{l}{\MyIndent US Data}         & 2.24          & 2.48          & 2.50         & 2.67          & 2.86          \\ \hline\hline
\multicolumn{1}{l}{RNN (Ours)} \\
\multicolumn{1}{l}{\MyIndent US Data}      & 3.46   & 3.37 & 3.01 & 3.23 & 3.30\\
\multicolumn{1}{l}{\MyIndent World Data}      & 2.35   & 2.52 & \textbf{2.50} & \textbf{2.62} & \textbf{2.60}\\ \hline
\multicolumn{1}{l}{AutoML (Ours)} \\
\multicolumn{1}{l}{\MyIndent US Data}      & 2.41   & 2.58 & 2.71 & \textbf{2.45} & 2.92\\
\multicolumn{1}{l}{\MyIndent World Data}      & \textbf{1.97}   & \textbf{2.32} & 2.59 & \textbf{2.62} & \textbf{2.61}\\ \hline\hline
\multicolumn{1}{l}{SPF Median}         & 1.86          & 2.11          & 2.36         & 2.46          & 2.65          \\ \hline\hline
\end{tabular}}
% \caption{This table shows the performance of the best, mean, and median forecasts for the performance of gated recurrent units on the test set 2009-Q1 to 2020-Q1. Baseline models are also shown.}
\end{table}

To illustrate the effect that pooling data has on forecasting, we show a graph that orders RMSE performance based on increasing model complexity with RMSE performance, comparing the trend when estimated on US data versus pooled data.  

\begin{figure}[htbp]
    \centering
    \includegraphics[width=10cm]{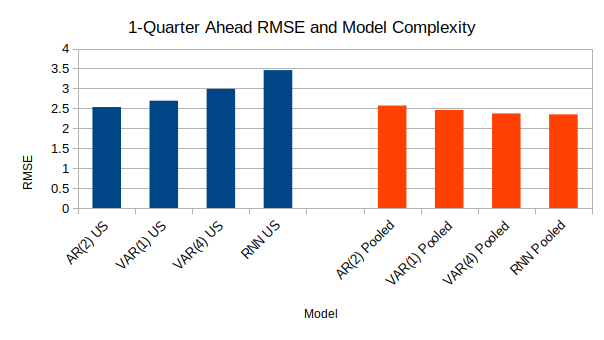}
    \caption{RMSE and Model Complexity}
    \label{fig:RMSE and model complexity}
\end{figure}

As the image shows, when using increasingly complex models with only the three hundred or so US timesteps, the most parsimonious model, the AR(2),performs the best and the models get progressively worse. However, when using the pooled data, the picture is entirely different. The AR(2) model actually performs a little bit worse -- likely due to chance. However, each progressively larger model improves on the AR(2) performance. Even if the RMSE decline is less striking in this latter case, this decline is never less compelling as the improvement is actually quite large but looks small as the US data RNN performs so poorly, it would never be used. In fact, despite the appearance of only a small improvement due to model complexity, the performance of the RNN on pooled data is state-of-the-art, while the performance of the AR(2) on pooled data is somewhat pedestrian. A similar story holds across other horizons with less striking consistency compared to the one period ahead story.  

We also provide a graph of the forecasts of five of the models: AR(2), factor, DSGE, RNN, and AutoML, as well as the true data. This graph is useful in disentangling why our machine learning models outperform. To illustrate the relative strengths of the models, we display the one quarter ahead forecasts in Figure \ref{fig:Forecasting Graph one quarter} here and the rest of the graphs are in Appendix \ref{graphs_appendix}.

\begin{figure}[htbp]
    \centering
    \includegraphics[width=10cm]{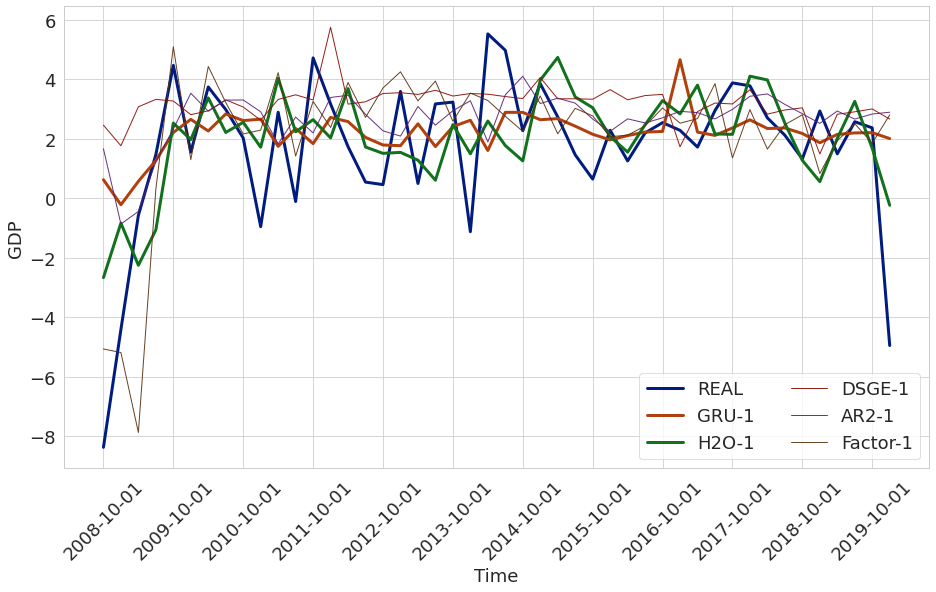}
    \caption{One Quarter Ahead - Forecasts}
    \label{fig:Forecasting Graph one quarter}
\end{figure}

AutoML, factor, and RNN models all did a good job at forecasting the Great Recession, with AutoML forecasting the best at one quarter ahead. The AR(2) and DSGE did not detect a regime change for the recessionary periods and are also upwardly biased leading to even worse performance during recession onset.\footnote{More information on the biases and variances of the different models can be found in Appendix \ref{bias_appendix}.} However, the XGBoost model that performed the best in AutoML was satisfactory forecast the expansions both in terms of the average level as well as individual movements in the quarterly data. Neither the factor model nor the RNN were able to forecast the quarter by quarter movements with such accuracy. More information about the performance, with a focus on our RNN model, is labeled in our robustness checks in Appendix \ref{robustness_appendix}. 

\section{Conclusion} \label{section_conclusion}
In this paper we show how estimating macroeconomic models on a panel of countries, as opposed to a single country can significantly improve external validity. Using a panel of countries as a training set, we statistically improved the RMSE performance of reduced-form models -- AR(2), VAR(1), and VAR(4) -- by roughly 12\%. We further show that we can make these reduced-form models more policy/country invariant, suggesting that these models have learned to generalize GDP forecasting even to countries the model has never been trained on. 

We also showed that a similar training set of a panel of countries can improve external validity of structural models which again are typically estimated only on a single country of interest. We focus on the Smets-Wouters model \citep{smets2007shocks}. Using a panel of countries improves the forecasting performance of the Smets-Wouters model estimated with maximum likelihood by roughly 24\% averaged across horizons. These results are again statistically significant. We then demonstrated that we can again improve policy-invariance and generalization to out-of-sample countries by using a panel of countries in our training set. Additionally, we addressed one potential roadblock to the adoption of pooling country data, which is the fact that the structural parameters may not be stable across countries and hence the pooled parameter value can only be interpreted as a mean value. While our results are less conclusive on this front, we argue that based on a forecasting exercise that parameter generalization and stability are likely as good across space as across time. Finally, concluding our section on structural models, we capitalize on the consistency of improvements and discuss the likelihood that our results will extend to other estimation techniques like generalized method of moments, calibration, and Bayesian approaches. 

Our last set of results, recognizes that our dataset has increased from 300 timesteps to around 3000 timestep-countries, showing that nonparametric machine learning models are able to outperform all the economic baseline models even after being estimated in this more data rich regime. Our RNN outperforms all economic baselines for horizons longer than two periods ahead. Likewise, our AutoML model outperforms all baselines for all horizons except for the three quarters ahead. Combined, the best performing model over all horizons is either an AutoML model or a recurrent neural network model which suggests there is likely much more room to test other nonparametric models in the more data rich macroeconomic regime.

\newpage

\bibliographystyle{aea.bst}
\bibliography{refs.bib}

\newpage
\section{Appendix}
\renewcommand{\thesubsection}{\Alph{subsection}}

\subsection{Selected Countries}
\label{countries_appendix}
%Egypt, India, Nigeria, Saudi Arabia, Ukraine
Countries in reduced-form data set: Australia, Austria, Belgium, Brazil, Canada, Switzerland, Chile, Columbia, Cyprus, Czech Republic, Germany, Denmark, Spain, Estonia, European Union, Finland, France, Great Britain, Greece, Hong Kong, Croatia, Hungry, Ireland, Israel, Italy, Japan, Korea, Luxembourg, Latvia, Mexico, Mauritius, Malaysia, Netherlands, Norway, New Zealand, Peru, Philippines, Poland, Portugal, Romania, Russia, Singapore, Slovakia, Slovenia, Sweden, Thailand, Turkey, USA and South Africa. 

Countries in structural data set: Australia, Austria, Belgium, Canada, Chile, Columbia, Germany, Denmark, Spain, Estonia, Finland, France, Iceland, Israel, Italy, Japan, Korea, Lithuania, Luxembourg, Mexico, Netherlands, New Zealand, Poland, Portugal, Slovakia, Slovenia, Sweden, USA 

\subsection{Selected Performance: Graphs}
\label{graphs_appendix}
\begin{figure}[htbp]
    \centering
    \includegraphics[width=10cm]{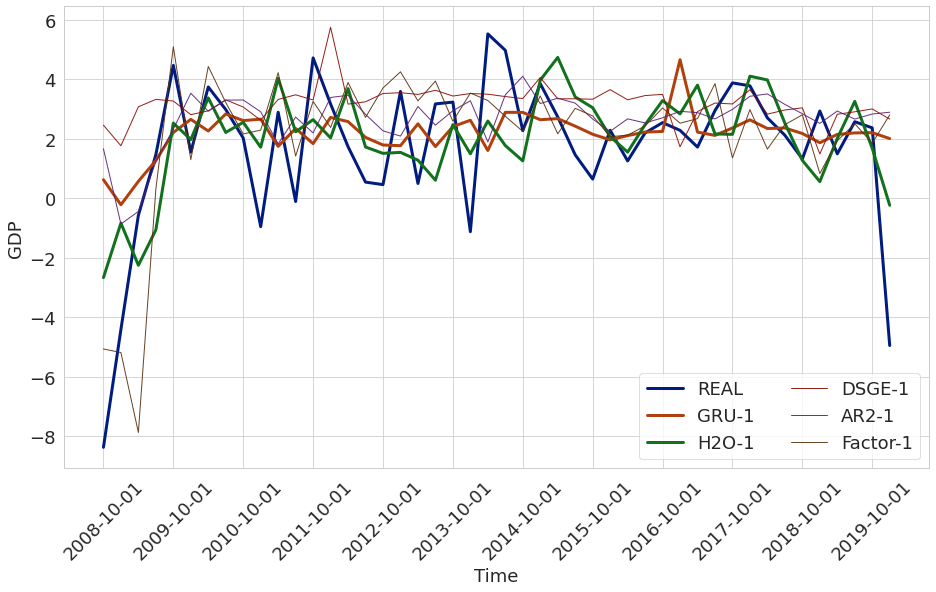}
    \caption{One Quarter Ahead - Forecasts}
    % \label{fig:Forecasting Graph1}
\end{figure}
\begin{figure}[htbp]
    \centering
    \includegraphics[width=10cm]{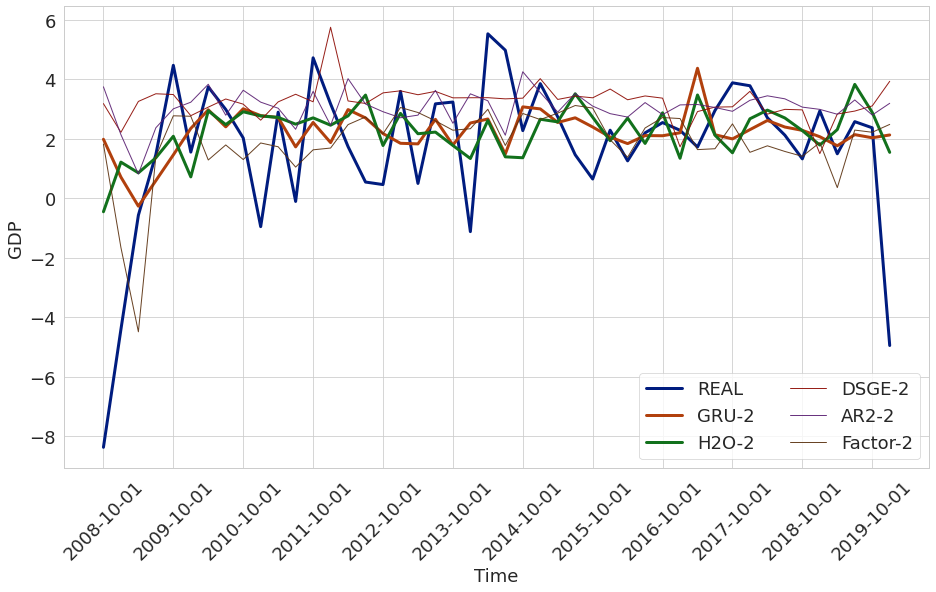}
    \caption{Two Quarters Ahead - Forecasts}
    % \label{fig:Forecasting Graph1}
\end{figure}
\begin{figure}[htbp]
    \centering
    \includegraphics[width=10cm]{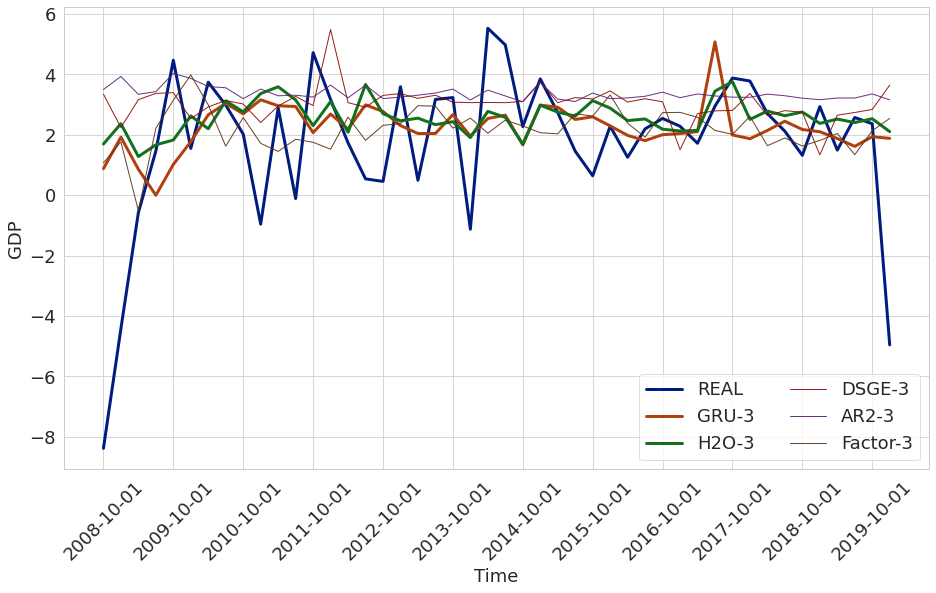}
    \caption{Three Quarters Ahead - Forecasts}
    % \label{fig:Forecasting Graph1}
\end{figure}
\begin{figure}[htbp]
    \centering
    \includegraphics[width=10cm]{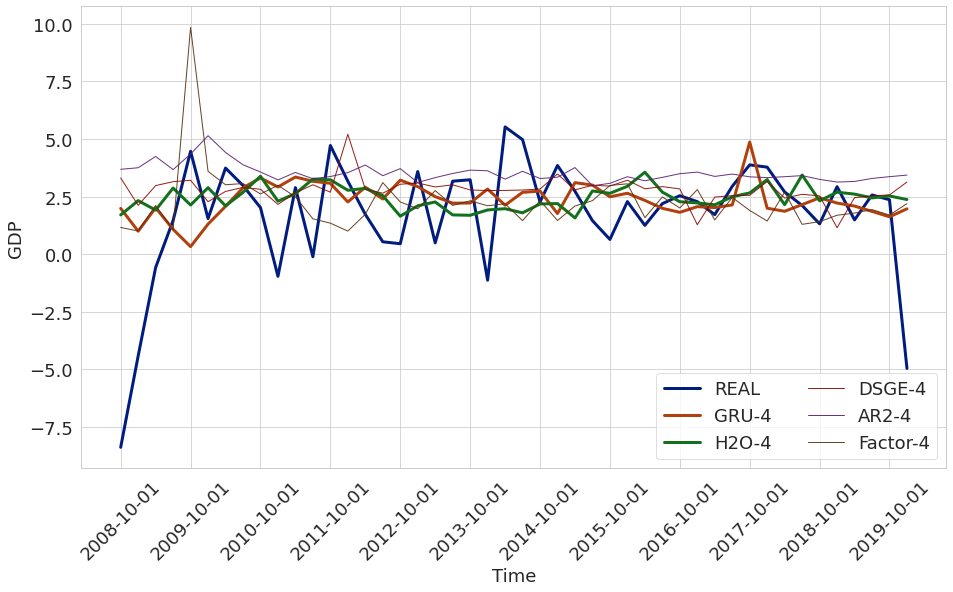}
    \caption{Four Quarters Ahead - Forecasts}
    % \label{fig:Forecasting Graph1}
\end{figure}
\begin{figure}[htbp]
    \centering
    \includegraphics[width=10cm]{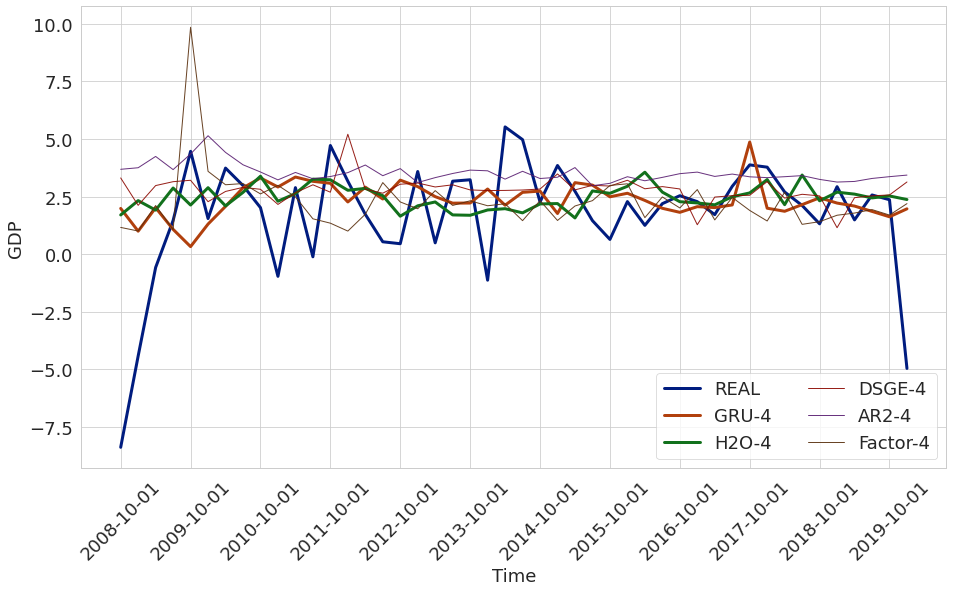}
    \caption{Five Quarters Ahead - Forecasts}
    % \label{fig:Forecasting Graph1}
\end{figure}

% \begin{figure}[htbp]
%     \centering
%     \includegraphics[width=10cm]{GRU-1_dw.png}
%     \caption{1-Quarter Ahead - Expansions Data}
%     \label{fig:Forecasting Graph}
% \end{figure}
% \begin{figure}[htbp]
%     \centering
%     \includegraphics[width=10cm]{GRU-2_dw.png}
%     \caption{2-Quarters Ahead - Expansions Data}
%     \label{fig:Forecasting Graph}
% \end{figure}
% \begin{figure}[htbp]
%     \centering
%     \includegraphics[width=10cm]{GRU-3_dw.png}
%     \caption{3-Quarters Ahead - Expansions Data}
%     \label{fig:Forecasting Graph}
% \end{figure}
% \begin{figure}[htbp]
%     \centering
%     \includegraphics[width=10cm]{GRU-4_Dw.png}
%     \caption{4-Quarters Ahead - Expansions Data}
%     \label{fig:Forecasting Graph}
% \end{figure}
% \begin{figure}[htbp]
%     \centering
%     \includegraphics[width=10cm]{GRU-5_dw.png}
%     \caption{5-Quarters Ahead - Expansions Data}
%     \label{fig:Forecasting Graph}
% \end{figure}

\pagebreak

\subsection{Details on the Survey of Professional Forecasters}
\label{spf_appendix}
While our model used the 2020 vintage of data, in reality, the forecasters for the Survey of Professional Forecasters were working with pseudo-out-of-sample vintages when forecasting over the entire test. While reproducing this would be possible from using old vintages, it would require estimating the model at every time step of the test set as the data would change every period. We wanted to avoid this pseudo-out-of-sample forecasting as it would result in estimating 4600 models instead of 100 at each horizon. Beyond this, the benefit of this increased computation was not clear as we would still be using 2020 vintage data for countries outside the US, many of which old vintages are difficult or impossible to find. So, it was easier to compare the SPF performance on the 2020 vintage. Plus, all our baseline models including world forecasts were estimated and evaluated using the 2020 vintage as well, so this choice allowed us to compare the SPF performance with the performance of all the baseline models. 

\subsection{Detailed Description of Economic Baseline Models}
\label{econ_models_appendix}
A first model we use is the autoregressive model, AR(n). An oft-used benchmark model, it estimates a linear relationship using the independent variable lagged N times. In terms of forecasting ability, this model is competitive with or outperforms the other economic models in our tests which is consistent with \citet{diebold1998elements}. We used an autoregressive model with two lags and a constant term. 

Additionally, we compared the Smets-Wouters 2007 model \citep{smets2007shocks}, as DSGE models share many similarities with recurrent neural networks and Smets-Wouters (2007) suggests that this particular model can outperform VARs and BVARs in forecasting. When running this, we used the standard Smets-Wouters Dynare code  contained on the published paper's data appendix. We take the point forecasts from the \citet{smets2007shocks} and use that to forecast. Like \citet{smets2007shocks}, we use Dynare \citep{Adjemianetal2011} to solve and estimate the model.  

A final model we included in our baseline economic models were factor models (see \citet{stock2002forecasting} and \citet{stock2002macroeconomic}). In short, the factor model approach takes a large cross-section of data and uses a technique like principal components analysis to reduce the dimensionality of the problem. In our case, we concatenate five to eight principal components based on information criterion of the high dimensional data with a lagged value of GDP and regress future GDP. We modified and used the code from FRED-QD as our baseline factor model \citep{mccracken2016fred}. While these models were extremely effective at lower horizons, these models were also dependent on a large cross-section of economic data with a long history in a country. In reality, only a few other developed countries have a cross-section of data that would be large enough to permit using these models as effectively as can be used in the United States. That being said, factor models do outperform our neural networks at shorter time intervals, and we imagine there is promise in combining the factor approach with a RNN or AutoML approach. 

We also tested the the forecasting performance of vector autoregressions \citep{10.2307/1912017}. In addition to displaying performance in our main table, we compared this model and the AR(2) in our 50 countries cross-section test as well. Since we were only forecasting GDP, the vector autoregressive models used lagged GDP, consumption, and unemployment used to forecast the single GDP variable as a linear regression. 

All the economic models were estimated on US GDP as is standard. While we ran preliminary tests on estimating these models on our cross-section of 50 countries, we ran into issues with estimating both factor models and DSGE models this way. However, preliminary results on the AR(2) model suggests there could be some improvement to using a cross-section even on a three parameter AR(2) model. The improvement is not as large as the RNN, which is not surprising as the RNN has more parameters to take advantage of a larger data set.

\subsection{Neural Network Models}
\subsubsection{Feed-Forward Neural Networks}
\begin{figure}[htbp]
    \centering
    \includegraphics[width=10cm]{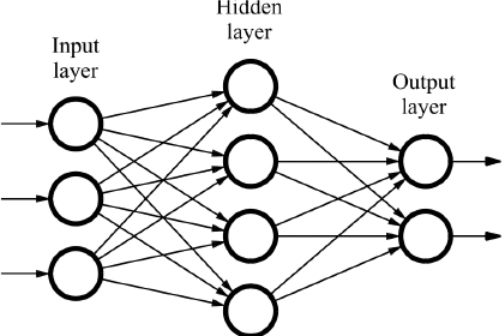}
    \caption{An Example of a Feed-Forward Neural Network}
     \label{fig:Feed-Forward Net}
\end{figure}

The feed-forward network is the prototypical image associated with deep learning. At its core, a feed-forward neural network is a recursively nested linear regression with nonlinear transforms. For example, assume $X$ is a vector valued input to the neural network and $X_{out}$ is the output. In a typical linear regression, $X_{out} = X_{in}\beta_{1}$. The insight for composing a feed-forward network is to take the output and feed that into another linear regression: $Y = X_{out}\beta_{2}$. In Figure \ref{fig:Feed-Forward Net}, $X_{in}$ would be the input layer, $X_{out}$ would be the hidden layer and $Y$ would be the output layer. The problem is not all that interesting if $X_{out}$ is a scalar. If $X_{in}$ is a matrix of dimension timesteps by regressors, $X_{out}$ can be a matrix of dimension timesteps by hidden units. Here in the figure, the dimension of the hidden layer is four, so $\beta_1$ has to be a matrix of dimension three by four (regressors by hidden units). Thus, we make $X_{out}$ an input into multidimensional regression for the second layer, $Y = X_{out}\beta_{2}$, if the first layer is a vector regression.\footnote{Note: this regression is not a vector autoregression as $X_{out}$ is a latent variable} This can be repeated for as many layers as desired. 

Now a composition of two layers will result in: $Y =X_{out}\beta_{2} = (X_{in}\beta_{1})\beta_{2}$. A product of two matrices is still another matrix, which means the model is still linear. Clearly this will hold no matter how many layers are added. However, an early result in the literature showed that if between every regression, eg $X_{out} = X_{in}\beta_{1}$, one inserts an almost arbitrary nonlinear link function this allows a neural network to approximate any continuous function \citep{HORNIK1989359}. For example, inserting a logistic transformation between $X_{in}$ and $X_{out}$ i.e. $X_{out} = \sigma (X_{in}\beta_{1})$ where $\sigma(z)= \frac{1}{1+e^{-z}}$. One can put these nonlinearities as often as one would like to get something like this: $Y = \sigma(\sigma (X_{in}\beta_{1,1})\beta_{2})$. These are the fundamental building blocks of neural networks and additional composition of link functions and matrix multiplication of parameters form the basis of deeper networks and allows these models to be universal approximators. 

\subsubsection{The Simplest Recurrent Neural Network: A Linear State Space Model}
For the purposes of this paper, associating an RNN with a feed-forward networks will not hinder comprehension. That being said, without even knowing it, many economists are already familiar with RNNs. The simplest is a Kalman filter-like linear state space model. The two equations that define the linear state space model are\footnote{We add autoregressive lags to make the model more general.}:
\label{eqn:eq1a,2a}
\begin{align}
\textbf{s}_{t} &= \textbf{s}_{t-1}\textbf{U}^s+\textbf{y}_{t-1}\textbf{W}^s+\textbf{b}^s , \\
\textbf{y}_{t} &= \textbf{s}_{t}\textbf{U}^y+\textbf{y}_{t-1}\textbf{W}^y+\textbf{b}^y
\end{align}

In a linear state space model, the state $\textbf{s}_i$ is an unobserved variable which allows the model to keep track of the current environment. One uses the state, along with lagged values of the observed variables, to forecast observed variables $\textbf{y}_i$. For example, for GDP, the state could be either an expansionary period or a recession -- a priori, the econometrician does not know. However, one can make an educated guess based on GDP growth. As machine learning is more interested in prediction, the state is often estimated with point estimates, which allows the data scientist to sidestep the tricky problem of filtering.

\subsubsection{Estimating the Parameters of a Linear State Space Model on Data}
The two equations that define the linear state space model are
\begin{align}
Y_{t} &= D*S_{t}+E*Y_{t-1}+F ,  \label{eqn:eq1}\\ 
S_{t} &= A*S_{t-1}+B*Y_{t-1}+C  \label{eqn:eq2}
\end{align}

We use Equations \ref{eqn:eq1} and \ref{eqn:eq2} to recursively substitute for the model prediction at a particular time period so the forecast for period 1 then is:

\begin{align}
\hat{y}_{1} =  D*(A*0+B*Y_{0}+C)+E*Y_{0} + F \label{eqn:eq3}
\end{align} 
and the forecast for period 2 is: 

\begin{align}
\hat{y}_{2} = D*(A*(A*0+B*Y_{0}+C)+B*Y_{1}+C)+E*Y_{1}+F \label{eqn:eq4}
\end{align}
Hatted variables indicate predictions and unhatted variables correspond to actual data. Additional time periods would be solved by iteratively substituting for the state using Equations \ref{eqn:eq1} and \ref{eqn:eq2} for the previous state. In order to update the parameters matrices $A, B, C, D, E,$ and $F$, the gradient is derived for each matrix and each parameter is updated via hill climbing. We will illustrate the process of hill climbing by taking the gradient of one parameter $B$:

\begin{align}
\frac{\partial \sum_{\forall t} L(y-\hat{y})}{\partial B} =  \frac{\partial L(y_{1}-\hat{y}_{1})}{\partial B} + \frac{\partial L(y_{2}-\hat{y}_{2})}{\partial B} \label{eqn:eq5}
\end{align}
Here $L()$ indicates the loss function. Substituting for $y'_{1}$ and $y'_{2}$ with Equations \ref{eqn:eq3} and \ref{eqn:eq4} into \ref{eqn:eq5} and using squared error as the loss function, we arrive at an equation with which we can take partial derivatives for with respect to A:

\begin{multline} \label{eqn:eq6}
\frac{\partial }{\partial B}L =  \frac{\partial}{\partial B}\frac{1}{2}(y_{1}-D*(A*0+B*Y_{0}+C)+E*Y_{0} + F)^2 \\
+ \frac{\partial}{\partial B}\frac{1}{2}(y_{2}-D*(A*(A*0+B*Y_{0}+C)+B*Y_{1}+C)+E*Y_{1}+F)^2 
\end{multline}

Distributing all the $B$'s and taking the derivative of \ref{eqn:eq6} results in 
$\frac{\partial }{\partial B}L = -(y_{1}-D*(A*0+B*Y_{0}+C)+E*Y_{0} + F)*D*Y_{0}-(y_{2}-D*(A*(A*0+B*Y_{0}+C)+B*Y_{1}+C)+E*Y_{1}+F)*(D*A*Y_{0} + D* Y_{1})$ which provide the gradients for hill climbing. In practice, the derivatives are taken automatically in code. 

\subsubsection{Gated Recurrent Units}
\label{gru_appendix}
Gated recurrent units \citep{DBLP:journals/corr/ChoMBB14} were introduced to improve upon the performance over previous RNNs that resembled linear state space models and can deal with the exploding gradient problem. 

The problem with linear state space models is that if one does not apply filtering, the state vector either blows up or goes to a steady state value. This can be seen by recognizing that each additional timestep results in the state vector getting multiplied by $\textbf{U}^s$ an additional time. Depending on if the eigenvectors of $\textbf{U}^s$ are greater than or less than one, the states will ultimately explode (go to infinity) or go to a steady state. More sophisticated RNNs like gated recurrent units \citep{DBLP:journals/corr/ChoMBB14} we use, fix this with the use of gates.

First, we redefine sigma as the logistic link function:
\label{eqn:eq7}
\begin{align}
\sigma (x) = \frac{e^{\beta x}}{1+e^{\beta x}} 
\end{align}
The idea behind the gate, is to allow the model to control the magnitude of the state vector. A simple gated recurrent neural network looks like the linear state space model with an added gate equation: 

%Todo issue here with alignment
\label{eqn:e7a}
\begin{align}
\textbf{y}_{t} = \textbf{h}_{t}\textbf{U}^y+E*\textbf{y}_{t-1}\textbf{W}^y+\textbf{b}^y \label{eqn_rnn1} \\
\textbf{z}_t =  \sigma(\textbf{h}_{t-1}\textbf{U}^h+\textbf{y}_{t-1}\textbf{W}^h+\textbf{b}^h) \label{eqn_rnn2} \\
\textbf{s}_{t} = \textbf{h}_{t-1}\textbf{U}^s+\textbf{y}_{t-1}\textbf{W}^s+\textbf{b}^s \label{eqn_rnn3} \\
\textbf{h}_t =  \textbf{z}_t \odot \textbf{s}_t \label{eqn_rnn4}
\end{align}

The output of $\sigma()$ is a number between zero and one which is element-wise multiplied by $\textbf{s}_t$, the first draft of the state. The operation $\odot$ indicates element-wise multiplication or the Hadamard product. Variables are subscripted with the time period they are observed in ($t$ or $t-1$). Weight matrices, which are not a function of the inputs, are superscripted with the equation name they feed into. All elements are considered vectors and matrices, and matrix multiplication is implied when no operation is present. 

The presence of the gate controls the behavior of the state, which means that even if the eigenvalues of $\textbf{U}^s$ were greater than one, or equivalently even if $\textbf{h}_t$ would explode without the gate, the gate can keep the state bounded. Additionally, the steady state distribution of the state does not have to converge to a number. The behavior could be periodic, or even chaotic \citep{zerroug2013chaotic}. This allows for the modeling of more complex behavior as well as the ability of the state vector to ``remember'' behavior over longer time periods \citep{chung2014empirical}. 

The equations of the gated recurrent unit are: 
\label{eqn:eq9}
\begin{align}
\textbf{y}_{t} = \textbf{h}_{t}\textbf{U}^y+E*\textbf{y}_{t-1}\textbf{W}^y+\textbf{b}^y \label{eqn_gru1} \\
\textbf{z}_t =\sigma(\textbf{x}_t\textbf{U}^z + \textbf{h}_{t-1} \textbf{W}^z) \label{eqn_gru2}  \\
\textbf{r}_t =\sigma(\textbf{x}_t \textbf{U}^r +\textbf{h}_{t-1} \textbf{W}^r) \label{eqn_gru3}  \\
\textbf{s}_t = \tanh(\textbf{x}_t \textbf{U}^s + (\textbf{h}_{t-1} \odot \textbf{r}_t) 
\textbf{W}^s) \label{eqn_gru4}  \\
\textbf{h}_t = (1 - \textbf{z}_t) \odot \textbf{s}_t + \textbf{z}_t \odot \textbf{h}_{t-1} \label{eqn_gru5} %this is the measurement equation 
\end{align}

Tanh is defined as the hyperbolic tangent:
\label{eqn:eq8}
\begin{align}
\tanh (x) = \frac{e^{2* x}-1}{e^{2* x}+1}
\end{align}

Like the linear state space model, the state vector of the gated recurrent unit persists over timesteps in the model. Mapping these equations to Equation \ref{eqn_rnn1}-\ref{eqn_rnn4}, Equation \ref{eqn_gru5} is the measurement equation (analogous to Equation \ref{eqn_rnn1}). Equation \ref{eqn_gru2} and \ref{eqn_gru3} are both gates and analogous to Equation \ref{eqn_rnn2}. Equation \ref{eqn_gru4} is the first draft of the state before the gate $\textbf{z}_t$ is applied and resembles Equation \ref{eqn_rnn3}. Equation \ref{eqn_gru5} is the final draft of the state after $\textbf{z}_t$ is applied and resembles Equation \ref{eqn_rnn4}.

The recurrent neural network is optimized using gradient descent, where the derivative of the loss function with respect to the parameters is calculated via the chain rule/reverse mode differentiation. The gradient descent optimizer algorithm we use is Adam \citep{Adam}, which shares similarities with a quasi-Newton approach. See Appendix \ref{adam_subsection} for more information.

\subsubsection{The Rectified Linear Unit}
\label{relu_appendix}
A nonlinearity used in our architecture, but not in the gated recurrent unit layers is the rectified linear unit (ReLU) \citep{DBLP:journals/corr/abs-1803-08375}. The rectified linear unit is defined as: 
\label{eqn:eq8.5}
\begin{align}
ReLU(x) = max(0,x)
\end{align}
The ReLU is the identity operation with a floor of zero much like the payoff of a call option. Despite being almost the identity map, this nonlinearity applied in a wide enough neural network can approximate any function \citep{HORNIK1989359}. 

\subsubsection{Skip Connections and Batch Norm}
\label{batch_norm_appendix}
Skip connections \citep{he2015deep} allow the input to skip the operation in a given layer. The input is then just added onto the output of the skipped layer, forming the final output of the layer. This allows the layer being skipped to learn a difference between the ``correct'' output and input, instead of learning a transformation. Additionally, if the model is overfitting, the neural network can learn the identity map easily. Skip connections are used when the input and the output are the same dimension which allows each input to correspond to one output. Because our network does not have this property, we learn a linear matrix that converts to the input to the output dimension. All the skip connections are linear operations and have no activation or batch norm, which differs from the pair of dense layers at the beginning of the network which have both batch norm and rectified linear unit activations. 

Batch normalizing \citep{ioffe2015batch} is used to prevent drift of output through a deep neural network. Changes to parameters in the early layers will cause an out-sized effect on the output values for the later layers. Batch norm fixes this problem by normalizing the output to look like a standard normal distribution after the output of each layer. Thus the effect of changes in parameters will not greatly effect the magnitude of output vector as between each output the data is re-normalized to have a mean of 0 and a standard deviation on 1. 

\subsubsection{Adam Optimizer}
\label{adam_subsection}
Adam combines momentum \citep{Momentum} -- a technique that uses recent history to smooth out swings orthonormal to the objective direction -- with RMSprop \citep{RMSprop} -- a technique used to adjust step size based on gradient volatility. 

Traditional gradient descent hill climbing updates the parameters with a single equation: 
\label{eqn:a1}
\begin{align}
\theta_t = \theta_{t-1}-\lambda*\nabla_\theta L_\theta(x,y)
\end{align}

Here $\nabla_\theta L_\theta(x,y)$ denotes taking the gradient of the loss with respect to $\theta$, the parameters of the model. For convenience, I will denote this term $g_t$. By subtracting the gradient multiplied by a small step size $\lambda$, this is moving the parameters theta in the direction the reduces the loss the most at $\theta_{t-1}$

If we wanted to use information from the second derivative to inform optimization, we can use Newton-Raphson instead:
\label{eqn:a2}
\begin{align}
\theta_t = \theta_{t-1}-H_t^{-1}*g_t
\end{align}

This uses the Hessian to determine an optimal step size based on steepness in the loss function. Typically, this approach is not used in deep learning as deep learning models typically have a large number of parameters and calculating the Hessian has a quadratic cost in the number of parameters and inverting also has a super-linear cost. However, there are quasi-Newton methods that attempt to approximate the Hessian to determine the step size without the high computational cost. Adam is similar to these methods. The equations that define Adam are as follows:

\begin{align}
\nu_t = \beta_1*\nu_{t-1} - (1-\beta_1) g_t \\
s_t = \beta_2*s_{t-1} - (1-\beta_2)*g_t^2 \\
\delta \theta_t = -\eta\frac{\nu_t}{\sqrt{s_t+\epsilon}}*g_t \\
\theta_{t+1} = \theta_t + \delta \theta_t
\end{align}

\begin{figure}[htbp]
    \centering
    \includegraphics[width=10cm]{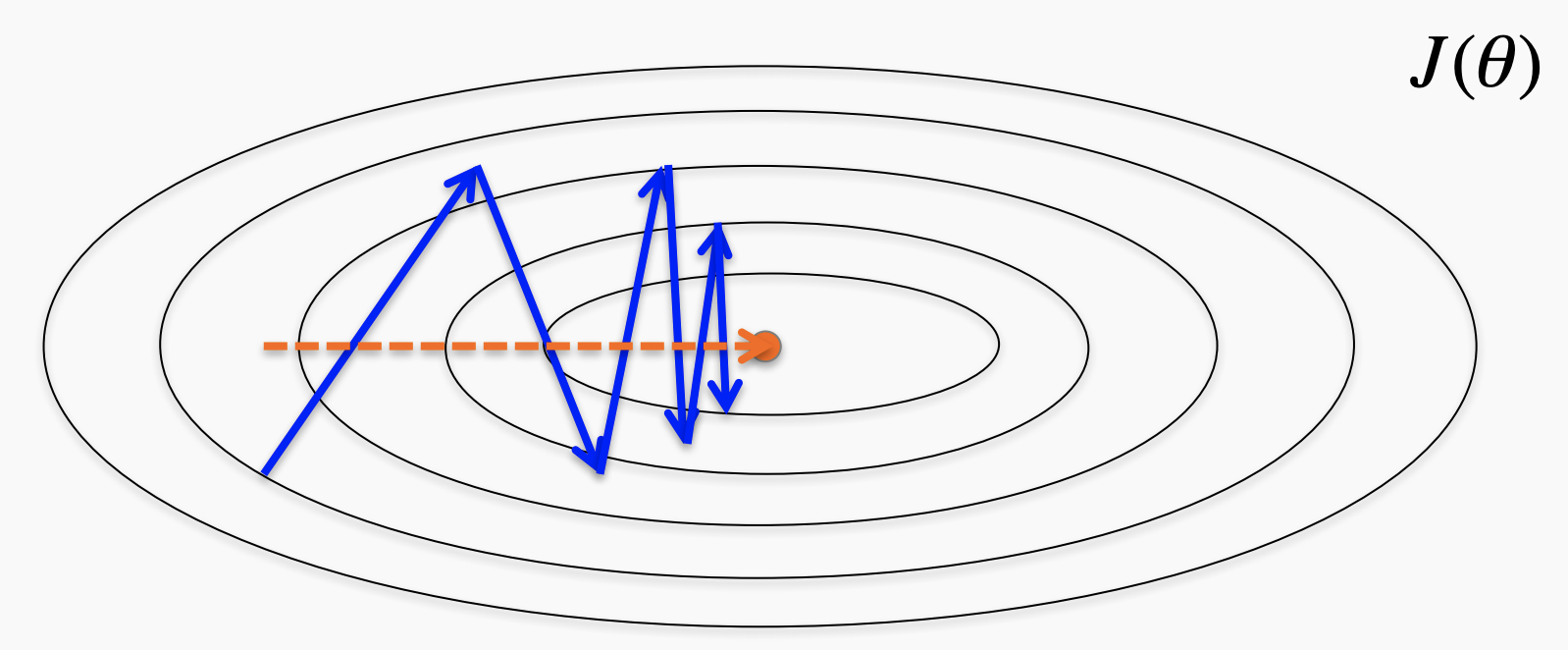}
    \caption{Momentum}
    \label{fig:Momentum figure}
\end{figure}

The first equation is a moving average of the gradient. This ``momentum'' term is used because often in training the direction of the gradient would move nearly perpendicular to the direction towards the optimum. Gradient descent would spend a lot of time zig-zagging while only making slow progress towards an optimum (see Figure \ref{fig:Momentum figure}). Taking a moving average of previous gradients preserves the principal direction while the orthogonal directions cancel each other out. 

Likewise, the $s_t$ equation is a moving average approximation for the Hessian. The approximate Hessian is used for adjusting the step size of the algorithm based on the curvature of the loss function at a given point. $\beta_1$ and $\beta_2$ are hyperparameters that determine the smoothness of the moving average. Again the resulting update term is applied to the previous values of the parameters. This approach is empirically shown to lead to more stable optimization and even better optima than simpler gradient descent approaches for large networks. 

\subsection{Additional Forecasting Information}
\subsubsection{Information Content Regressions}
\label{Information_Content_Regressions_appendix}
We regress true GDP on a varying collection of forecasts to test for statistically significant contribution of a given forecast like our gated recurrent unit model. An interpretation of significant coefficients would be that the given forecast method is adding statistically significant information when pooled with the other regressions. 

\newpage 

Here is the H2O forecast compared to the SPF on the baseline test set ranging from one-quarter ahead forecasts to five-quarters ahead:
% Table created by stargazer v.5.2.2 by Marek Hlavac, Harvard University. E-mail: hlavac at fas.harvard.edu
% Date and time: Tue, Nov 17, 2020 - 03:46:24 PM
\begin{table}[!htbp] \centering 
  \caption{} 
  \label{} 
\begin{tabular}{@{\extracolsep{5pt}}lccccc} 
\\[-1.8ex]\hline 
\hline \\[-1.8ex] 
\\[-1.8ex] & \multicolumn{5}{c}{Real GDP Growth} \\ 
\\[-1.8ex] & (1-Qtr) & (2-Qtrs) & (3-Qtrs) & (4-Qtrs) & (5-Qtrs)\\ 
\hline \\[-1.8ex] 
SPF & 0.796$^{***}$ & 1.554$^{***}$ & 3.042$^{***}$ & 2.888$^{***}$ & 1.218 \\ 
  & (0.238) & (0.346) & (0.607) & (0.742) & (1.025) \\ 
  & & & & & \\ 
 H2O & 0.505$^{**}$ & 0.564 & 0.511 & 0.213 & 0.337 \\ 
  & (0.236) & (0.399) & (0.571) & (0.669) & (0.666) \\ 
  & & & & & \\ 
\textit{N} & 46 & 46 & 46 & 46 & 46 \\ 
R$^{2}$ & 0.532 & 0.474 & 0.409 & 0.271 & 0.033 \\ 
Adjusted R$^{2}$ & 0.510 & 0.449 & 0.382 & 0.237 & $-$0.012 \\ 
Residual Std. Error (df = 43) & 1.790 & 1.898 & 2.012 & 2.235 & 2.573 \\ 
F Statistic (df = 2; 43) & 24.464$^{***}$ & 19.366$^{***}$ & 14.879$^{***}$ & 7.997$^{***}$ & 0.743 \\ 
\hline 
\hline \\[-1.8ex] 
\textit{Notes:} & \multicolumn{5}{r}{$^{***}$Significant at the 1 percent level.} \\ 
 & \multicolumn{5}{r}{$^{**}$Significant at the 5 percent level.} \\ 
 & \multicolumn{5}{r}{$^{*}$Significant at the 10 percent level.} \\ 
\end{tabular} 
\end{table} 

\newpage

The following table contains regressions comparing the information content of the H2O and baselines, excluding the SPF: 
% Table created by stargazer v.5.2.2 by Marek Hlavac, Harvard University. E-mail: hlavac at fas.harvard.edu
% Date and time: Tue, Nov 17, 2020 - 03:59:28 PM
\begin{table}[!htbp] \centering 
  \caption{} 
  \label{} 
\begin{tabular}{@{\extracolsep{5pt}}lccccc} 
\\[-1.8ex]\hline 
\hline \\[-1.8ex] 
\\[-1.8ex] & \multicolumn{5}{c}{Real GDP Growth} \\ 
\\[-1.8ex] & (1-Qtr) & (2-Qtrs) & (3-Qtrs) & (4-Qtrs) & (5-Qtrs)\\ 
\hline \\[-1.8ex] 
  H2O & 0.778$^{***}$ & 1.385$^{***}$ & 1.198$^{*}$ & 0.665 & 0.151 \\ 
  & (0.259) & (0.427) & (0.691) & (0.779) & (0.684) \\ 
  & & & & & \\ 
 DSGE & 0.724 & $-$0.041 & 0.138 & 0.047 & 0.081 \\ 
  & (0.478) & (0.581) & (0.647) & (0.679) & (0.712) \\ 
  & & & & & \\
 AR2 & $-$0.554 & $-$1.102 & $-$0.981 & $-$0.797 & $-$1.576 \\ 
  & (0.438) & (0.781) & (1.759) & (1.088) & (1.634) \\ 
  & & & & & \\ 
 Factor & 0.358$^{*}$ & 0.657$^{*}$ & 0.898 & 0.501 & 0.506 \\ 
  & (0.188) & (0.354) & (0.539) & (0.323) & (0.463) \\ 
  & & & & & \\ 
\textit{N} & 46 & 46 & 46 & 46 & 46 \\ 
R$^{2}$ & 0.501 & 0.291 & 0.127 & 0.070 & 0.031 \\ 
Adjusted R$^{2}$ & 0.453 & 0.222 & 0.041 & $-$0.021 & $-$0.063 \\ 
Residual Std. Error (df = 41) & 1.893 & 2.257 & 2.505 & 2.586 & 2.638 \\ 
F Statistic (df = 4; 41) & 10.307$^{***}$ & 4.202$^{***}$ & 1.485 & 0.767 & 0.329 \\ 
\hline 
\hline \\[-1.8ex] 
\textit{Notes:} & \multicolumn{5}{r}{$^{***}$Significant at the 1 percent level.} \\ 
 & \multicolumn{5}{r}{$^{**}$Significant at the 5 percent level.} \\ 
 & \multicolumn{5}{r}{$^{*}$Significant at the 10 percent level.} \\ 
\end{tabular} 
\end{table} 

\newpage

This final table performs the same regression but has the SPF, H2O, and all baseline models:
% Table created by stargazer v.5.2.2 by Marek Hlavac, Harvard University. E-mail: hlavac at fas.harvard.edu
% Date and time: Tue, Nov 17, 2020 - 04:16:09 PM
\begin{table}[!htbp] \centering 
  \caption{} 
  \label{} 
\begin{tabular}{@{\extracolsep{5pt}}lccccc} 
\\[-1.8ex]\hline 
\hline \\[-1.8ex] 
\\[-1.8ex] & \multicolumn{5}{c}{REAL} \\ 
\\[-1.8ex] & (1-Qtr) & (2-Qtrs) & (3-Qtrs) & (4-Qtrs) & (5-Qtrs)\\  
\hline \\[-1.8ex] 
 H2O & 0.614$^{**}$ & 0.629 & 0.495 & 0.233 & 0.254 \\ 
  & (0.239) & (0.403) & (0.594) & (0.693) & (0.689) \\ 
  & & & & & \\ 
 SPF & 1.071$^{***}$ & 1.648$^{***}$ & 3.085$^{***}$ & 2.835$^{***}$ & 1.256\\ 
  & (0.326) & (0.394) & (0.693) & (0.776) & (1.145) \\ 
  & & & & & \\ 
 DSGE & 0.544 & $-$0.141 & $-$0.220 & $-$0.332 & $-$0.123 \\ 
  & (0.433) & (0.492)  & (0.542) & (0.605) & (0.734) \\ 
  & & & & & \\ 
 AR(2) & $-$1.029$^{**}$ & $-$0.981 & 0.776 & $-$0.749 & $-$1.672 \\ 
  & (0.420) & (0.660) & (1.509) & (0.954)& (1.632)  \\ 
  & & & & & \\ 
 Factor & 0.003 & 0.115 & 0.101 & 0.321 & 0.418 \\ 
  & (0.201) & (0.326) & (0.481) & (0.287) & (0.469)  \\ 
  & & & & & \\ 
\textit{N} & 46 & 46 & 46 & 46 & 46 \\ 
R$^{2}$ & 0.607 & 0.506 & 0.416 & 0.302 & 0.059 \\ 
Adjusted R$^{2}$ & 0.558 & 0.445 & 0.343 & 0.215 & $-$0.058 \\ 
Residual Std. Error (df = 40) & 1.701 & 1.907 & 2.074 & 2.267 & 2.632 \\ 
F Statistic (df = 5; 40) & 12.363$^{***}$ & 8.208$^{***}$ & 5.697$^{***}$ & 3.467$^{**}$ & 0.505 \\ 
\hline 
\hline \\[-1.8ex] 
\textit{Notes:} & \multicolumn{5}{r}{$^{***}$Significant at the 1 percent level.} \\ 
 & \multicolumn{5}{r}{$^{**}$Significant at the 5 percent level.} \\ 
 & \multicolumn{5}{r}{$^{*}$Significant at the 10 percent level.} \\ 
\end{tabular} 
\end{table} 

None of the models except the SPF have consistent statistically significant information above and beyond the other models. 

\newpage

\subsubsection{Bias and Variance in the Forecasting Models}
\label{bias_appendix}

The following table contains the mean bias as well as the variance of the models. For the gated recurrent unit we use the median forecast:

\begin{table}[!htbp] \centering 
  \caption{} 
  \label{} 
\begin{tabular}{@{\extracolsep{5pt}}lccccc} 
\\[-1.8ex]\hline 
\hline \\[-1.8ex] 
\\[-1.8ex] & \multicolumn{5}{c}{Forecast Bias} \\ 
\\[-1.8ex] & (1-Qtr) & (2-Qtrs) & (3-Qtrs) & (4-Qtrs) & (5-Qtrs)\\ 
\hline \\[-1.8ex] 
 RNN Bias & 0.459 & 0.480 & 0.506 & 0.620 & 0.644\\ 
  Variance & 5.51 & 6.34 & 6.23 & 6.85 & 6.75\\ 
  & & & & & \\ 
 H2O Bias & 0.293 & 0.511 & 0.833 & 0.723 & 0.422 \\ 
  Variance & 3.86 & 5.36 & 6.70 & 6.88 & 6.80 \\ 
  & & & & & \\ 
 SPF Bias & 0.331 & 0.600 & 0.723 & 0.804 & 0.901 \\
  Variance & 3.48 & 4.46 & 5.57 & 6.07 & 7.04 \\ 
  & & & & & \\ 
 DSGE Bias & 1.75 & 1.93 & 1.88 & 1.78 & 1.65 \\ 
  Variance & 9.32 & 10.77 & 10.76 & 10.42 & 9.99 \\ 
  & & & & & \\ 
 AR2 Bias & 0.404 & 0.389 & 0.431 & 0.472 & 0.481 \\ 
  Variance & 6.61 & 6.88 & 7.12 & 7.40 & 7.41 \\ 
  & & & & & \\
 VAR4 Bias & 0.233 & 0.214 & 0.201 & 0.200 & 0.195 \\ 
  Variance & 5.63 & 6.36 & 6.56 & 6.89 & 6.91 \\ 
  & & & & & \\ 
 Factor Bias & 0.432 & 0.163 & 0.459 & 0.533 & 0.699 \\ 
  Variance & 5.03 & 6.17 & 6.26 & 7.12 & 8.19 \\ 
  & & & & & \\ 
\hline 
\hline \\[-1.8ex] 
\end{tabular} 
\end{table} 

\newpage

\subsubsection{Parameter Comparison across US DSGE Model and World DSGE Model}
\label{param_comp}
The table shows the parameters of the DSGE model estimated on world data versus US data as well as the standard deviations of the parameters over time. This illustrates the important parameters that change when adding in global data. In addition to the model growing less confident and more accurate when world data is added, the parameters that are most modified are parameters governing wage stickiness and inflation. 

\newgeometry{left=.1cm, bottom=0.1cm}
\begin{table}[]
\scalebox{0.85}{
\begin{tabular}{llllll}
Dynare Variable            & Variable Description                                                       & World Param Values & US Param Values & World Standard Deviation & US Standard Deviation \\
'ea'        & Factor Productivity Shock Error                       & 0.976842           & 0.474861        & 0.034192                 & 0.014886              \\
'eb'        & Risk Permium Shock Error                              & 0.522046           & 0.260422        & 0.022821                 & 0.115139              \\
'eg'        & Government Spending Shock Error                       & 1.002773           & 0.659571        & 0.028303                 & 0.017909              \\
'eqs'       & Technology Shock Error                                & 1.377699           & 0.426465        & 0.114479                 & 0.033718              \\
'em'        & Monetary Policy Shock Error                           & 0.202341           & 0.209711        & 0.018569                 & 0.005405              \\
'epinf'     & Inflation Shock Error                                 & 0.785736           & 0.184302        & 0.034779                 & 0.019632              \\
'ew'        & Wage Shock Error                                      & 0.595469           & 0.316343        & 0.107472                 & 0.023861              \\
'crhoa'     & AR Parameter on productivity Shock        & 0.997478           & 0.984393        & 0.002449                 & 0.003019              \\
'crhob'     & AR Parameter on Risk Premium Shock        & 0.132395           & 0.383984        & 0.039344                 & 0.332214              \\
'crhog'     & AR Parameter on Government Shock          & 0.984414           & 0.980403        & 0.009976                 & 0.009258              \\
'crhoqs'    & AR Parameter on Technology Shock          & 0.688845           & 0.776188        & 0.109162                 & 0.027836              \\
'crhoms'    & AR Parameter on Monetary Shock            & 0.30509            & 0.148434        & 0.02064                  & 0.041207              \\
crhopinf'   & AR Parameter on Inflation Shock           & 0.611081           & 0.977832        & 0.032108                 & 0.024071              \\
'crhow'     & AR Parameter on Wage Shock                & 0.982497           & 0.899716        & 0.010334                 & 0.045943              \\
'cmap'      & AR Moving Average Error Term on Inflation & 0.539008           & 0.875734        & 0.038097                 & 0.058007              \\
'cmaw'      & AR Moving Average Error Term on Wages     & 0.970411           & 0.882725        & 0.011541                 & 0.046395              \\
'csadjcost' & Elasticity of the Capital Adjustment Cost             & 9.80101            & 8.15361         & 0.8643                   & 1.03545               \\
'csigma'    & Elasticity of Subsitution                             & 1.978862           & 1.895245        & 0.166713                 & 0.253334              \\
'chabb'     & Habbit Formation                                      & 0.868449           & 0.663259        & 0.016705                 & 0.173953              \\
'cprobw'    & Wage Flexibility Probability                          & 0.930401           & 0.924096        & 0.028501                 & 0.031956              \\
'csigl'     & Wage Elasticity of Labor Supply                       & 1.749702           & 3.763315        & 0.471358                 & 1.127552              \\
'cprobp'    & Price Flexibility Probability                         & 0.945883           & 0.663557        & 0.005239                 & 0.034912              \\
'cindw'     & Wage Indexation                                       & 0.01               & 0.635949        & 5.26E-18                 & 0.14763               \\
'cindp'     & Indexation to Past Inflation                          & 0.01               & 0.13723         & 5.26E-18                 & 0.123999              \\
'czcap'     & Elasticity of Capital Utilization                     & 0.364196           & 0.812247        & 0.194422                 & 0.080674              \\
'cfc'       & Fixed Costs in Production                             & 2.040895           & 1.93295         & 0.0107                   & 0.141187              \\
'crpi'      & Taylor Rule Inflation                  & 1                  & 2.114895        & 0                        & 0.321337              \\
'crr'       & Taylor Rule Interest Rate Smoothing    & 0.966891           & 0.912953        & 0.011906                 & 0.016476              \\
'cry'       & Taylor Rule Output Gap                   & 0.315889           & 0.118081        & 0.110856                 & 0.063766              \\
'crdy'      & Taylor Rule Output Gap Change            & 0.024124           & 0.15335         & 0.006121                 & 0.030106              \\
constepinf' & Gap between Model and Observed Inflation              & 0.187268           & 1.240916        & 0.023902                 & 0.2128                \\
'constelab' & Gap between Model and Observed Labor                  & 2.993805           & 0.640214        & 0.998992                 & 1.963757              \\
'ctrend'    & Gap between Model and Observed Output                 & 0.34744            & 0.426019        & 0.038315                 & 0.026315              \\
'cgy'       & Productivity Shocks on Government Spending  & 0.480685           & 0.550493        & 0.058911                 & 0.01411               \\
'calfa'     & Elasticity of Capital in Production Function          & 0.091021           & 0.212888        & 0.015633                 & 0.02167              
\end{tabular}}
\end{table}
\restoregeometry

\newpage

\subsection{Selected Additional Information}

\subsubsection{Smets-Wouters Model: US Data vs. World Data}
\label{smets_appendix}
\begin{table}[htbp] \centering
\caption{Smets-Wouters Root Mean Squared Error}
\caption*{The performance forecasts for the performance of Smets-Wouters model on the test set 2009-Q1 to 2020-Q1 (Lower is better)}
%uncertainty of multiple runs factored in the significant figures
\label{RMSE_Reduced_Form_Table}
\begin{tabular}{cccccc}
\hline \hline
% \multicolumn{6}{ c }{Root Mean Squared Error}\\
\multicolumn{1}{r}{Time (Q's Ahead)}   & 1Q           & 2Q           & 3Q           & 4Q           & 5Q           \\ \hline
\multicolumn{1}{l}{Smets-Wouters DSGE Max Like} \\
\multicolumn{1}{l}{\MyIndent US Data}         & 3.83          & 4.36          & 4.49         & 4.50          & 4.2          \\
\multicolumn{1}{l}{\MyIndent World Data}         & 3.05***          & 3.28***          & 3.28***         & 3.22**          & 3.16*          \\
\multicolumn{1}{l}{\MyIndent Out-Of-Sample Data}         & 2.59***          & 2.75***          & 2.89**         & 3.04          & 3.18          \\
\multicolumn{1}{l}{\MyIndent 1995+}         & 2.77***          & 3.09**          & 3.22*         & 3.26          & 3.28\\\hline
\multicolumn{1}{l}{Smets-Wouters DSGE Bayesian} \\
\multicolumn{1}{l}{\MyIndent US Data}         & 2.79          & 2.95         & 2.89         & 2.80          & 2.71          \\\hline\hline
\\\multicolumn{6}{r}{$^{*}$ Significance indicates outperformance of world data models over US data models} \\ 
\end{tabular}
\end{table}

\subsubsection{Recurrent Neural Network Robustness Checks}
\label{robustness_appendix}
For our RNN model, we found we could improve forecasting performance by taking the mean prediction of many models estimated by stochastic gradient descent. The ensembling improves performance slightly, but later graphs will show it also improves model stability and variance. Bolded entries indicate outperformance over all economic models. 

\label{ensemble_table}
\begin{table}[htbp] \centering
%uncertainty of multiple runs factored in the significant figures
%
\caption{RNN Mean, Median and Best forecasts}
\begin{tabular}{cccccc}
\hline \hline
% \multicolumn{6}{ c }{Root Mean Squared Error}\\
\multicolumn{1}{r}{Time (Q's Ahead)}   & 1Q           & 2Q           & 3Q           & 4Q           & 5Q           \\ \hline
\multicolumn{1}{l}{RNN with World Data} \\
\multicolumn{1}{l}{\MyIndent Best Model}      & 2.4   & 2.5 & \textbf{2.5} & \textbf{2.6} & \textbf{2.6}\\
\multicolumn{1}{l}{\MyIndent Mean Model}      & 2.3   & 2.5 & \textbf{2.5} & \textbf{2.6} & \textbf{2.6}\\
\multicolumn{1}{l}{\MyIndent Median Model}      & 2.3   & 2.5 & \textbf{2.5} & \textbf{2.6} & \textbf{2.6}\\\hline\hline
\end{tabular}
%\caption{This table shows the performance of the best, mean, and median forecasts for the performance of gated recurrent units on the test set 2009-Q1 to 2020-Q1. Baseline models are also shown}
\end{table}

We provide a Monte Carlo simulation (Table \ref{monte_carlo_table}), estimating our RNN model at each time horizon 100 times. At every horizon, the average root mean squared error of our simulated models indicates competitive, if not outperformance, against baseline models. Interestingly, it seems like the best performing model on validation data, when tested on the test data, often performs worse than the average performance over all the models. This is something that should be investigated further, but based on this phenomenon, we recommend that practitioners take a simple mean or median forecast across many different models. 

% \begin{center}
\begin{table}[htbp] \centering
\caption{Baseline RNN Monte Carlo Simulation over Initializations}
\caption*{The mean and standard deviation of the performance of gated recurrent units on the test set 2009-Q1 to 2020-Q1}
\label{monte_carlo_table}
\begin{tabular}{cccccc}
\hline \hline
\multicolumn{1}{r}{Time (Q's Ahead)}        & 1Q   & 2Q   & 3Q   & 4Q   & 5Q   \\ \hline
\multicolumn{1}{l}{Mean RMSE}              & 2.4   & 2.6  & 2.5  & \textbf{2.6}  & \textbf{2.6*} \\
\multicolumn{1}{l}{Std Dev RMSE}           & 0.06   & 0.06 & 0.05 & 0.06 & 0.06 \\ \hline \hline
\end{tabular}
%\caption{This table shows the mean and standard deviation of the performance of gated recurrent units on the test set 2009-Q1 to 2020-Q1. Baseline models performance are the same as in Table \ref{rmse_table}}
\end{table}
% \end{center}

A common criticism of deep learning attempts at forecasting is that the models are unreliable, but due to high variance, one can p-hack a model that performs well. The Monte Carlo simulation in Table \ref{monte_carlo_table}, anticipates this critique. The standard deviation of our RMSE is 0.06 which suggests that all our models have a similar performance on the test set when optimization is complete. This cannot resolve all issues, as the Monte Carlo result only deals with numerical instability. The model could still fit this particular data window or architecture choice, due to chance. In order to respond to those critiques, we also provide robustness checks across different architectures and data periods. 

One test we preformed was to replace the gated recurrent units with a long short-term memory (LSTM) layer \citep{Hochreiter:1997:LSM:1246443.1246450} -- another type of RNN. We use the same test data as the main result (USA 2009-Q1 to 2020-Q1) as well as the same data as inputs. The LSTM in Table \ref{monte_carlo_table_b} are analogous to the gated recurrent unit neural networks models in the table in Section \ref{nonpar_ml_models_section} in the main text. Mean RMSE and Std Dev RMSE, correspond to the entries in the table below. The baseline performances are still the same as the test set has not changed. 

% \begin{center}
\begin{table}[htbp] \centering
% \centering
\caption{Baseline LSTM Monte Carlo Simulations}
\caption*{The performance of the best, mean, and median forecasts as well as the mean and standard deviation of the long short-term memory networks on the test set 2009-Q1 to 2020-Q1 (lower is better)}
%combine this table with the original table 
\label{monte_carlo_table_b}
\begin{tabular}{cccccc}
\hline\hline
% \multicolumn{6}{ c }{RNN Monte Carlo Simulations}      \\
\multicolumn{1}{r}{Time (Q's Ahead)}        & 1Q   & 2Q   & 3Q   & 4Q   & 5Q   \\ \hline
\multicolumn{1}{l}{Best RMSE}              & 2.4   & 2.6  & 2.6  & \textbf{2.6}  & \textbf{2.6*} \\
\multicolumn{1}{l}{RMSE of Mean}              & 2.4   & 2.6  & 2.5  & \textbf{2.6}  & \textbf{2.6*} \\
\multicolumn{1}{l}{RMSE of Median}              & 2.4   & 2.5  & 2.5  & \textbf{2.6}  & \textbf{2.6*} \\
\multicolumn{1}{l}{Mean RMSE}              & 2.4   & 2.6  & 2.5  & \textbf{2.6}  & \textbf{2.6*} \\ 
\multicolumn{1}{l}{Std Dev RMSE}           & 0.05   & 0.07 & 0.05 & 0.06 & 0.06 \\ 
\hline\hline
\end{tabular}
%\caption{This table shows the performance of the best, mean, and median forecasts as well as the mean and standard deviation of the performance of Long Short-Term Memory networks on the test set 2009-Q1 to 2020-Q1. The baseline performance is the same as in table \ref{rmse_table}}
\end{table}
% \end{center}

The LSTM networks outperform the baseline models along essentially the same time horizons. Performance is also competitive, but consistently a little worse than the gated recurrent unit over all time frames. The LSTM has similar standard deviation of root mean squared error, suggesting that the two models consistently find a similar optimum when it comes to forecasting. Again, taking a model average through the mean or median forecast results in small but consistent root mean squared error performance improvements.

Additionally, we re-estimated the model with slightly different test set from 2009-Q4 to 2019-Q4 as opposed to 2008-Q4 to 2020-Q1, comparing the benchmark economic models to our original gated recurrent unit (Table \ref{rmse_table_a}). The reason we use this horizon is that it contains no recessions. Since the highly flexible neural network will have an advantage forecasting periods with a significant departure from a more linear-friendly period of expansion, removing the recessions would hamstring our model compared to the more linear model baselines. 

Our gated recurrent units were completely re-estimated as we additionally included 2009-Q1 to 2009-Q3 in the validation set. Performance would improve if we left those (recession) timesteps out of the validation set as the test set contains no recessions, however, this decision cannot be rationalized from the point-of-view of an out-of-sample forecaster. Although this version of our model did not outperform the best baseline models along any horizon, considering performance over all horizons, we think our median and mean models are better than the US AR(2), VAR(1), and the factor model on this test set, while performing slightly worse than the DSGE model and the world AR(2). This supports our hypothesis that the main outperformance of our model was in highly nonlinear domains like recessions and other regime changes although using the cross-sectional data reduced the tendency for the models to be biased upwards and was a contributor to the RNN's outperformance over models trained only on US data. 

% \begin{center}
% \begin{tabular}[c]{ c  c  c  c  c  c}
% \multicolumn{5}{ c }{Root Mean Squared Error}\\
%  Time (Q's Ahead) & 1Q & 2Q & 3Q & 4Q & 5Q\\ 
%  BVAR Litterman & 2.9 & 3.5 & 3.8 & 3.7 & 4.0 \\
%  VAR & 3.0 & 3.6 & 3.9 & 3.9 & 4.1\\
%  Smets Wouters DSGE & 2.6 & 2.7 & 2.7 & 2.7 & 2.7\\
%  SPF Median & 1.8 & 2.1 & 2.4 & 2.5 & 2.7\\
%  Our RNN Model & \textbf{1.8} & \textbf{1.9} & \textbf{2.0} & \textbf{2.2} & \textbf{2.2}\\
% \end{tabular}
\begin{table}[htbp]\centering
\caption{Expansion Root Mean Squared Error}
\caption*{Performance of the Best, Mean, and Median forecasts for gated recurrent units on the test set 2009-Q4 to 2019-Q4}
%one way to unify all the tables to show robustness of table 
\label{rmse_table_a}
\begin{tabular}{cccccc}
\hline\hline
% \multicolumn{6}{ c }{Root Mean Squared Error}\\
\multicolumn{1}{r}{Time (Q's Ahead)}   & 1Q           & 2Q           & 3Q           & 4Q           & 5Q           \\ \hline
\multicolumn{1}{l}{VAR(1)} \\
\multicolumn{1}{l}{\MyIndent US data}      & 2.3    & 2.6 & 2.9 & 3.0 & 3.0          \\ 
\multicolumn{1}{l}{\MyIndent World data}      & 2.1    & 2.2 & 2.2 & 2.2 & 2.2          \\ \hline
\multicolumn{1}{l}{AR(2)} \\
\multicolumn{1}{l}{\MyIndent US data}      & 1.7    & 1.7 & 1.8 & 1.9 & 1.9          \\ 
\multicolumn{1}{l}{\MyIndent World data}      & \textbf{1.6}    & 1.6 & 1.6 & \textbf{1.5} & \textbf{1.5*}          \\ \hline
\multicolumn{1}{l}{Smets Wouters DSGE} \\
\multicolumn{1}{l}{\MyIndent US data} & 1.8          & 1.8          & 1.7          & 1.6          & 1.5*         \\ \hline
\multicolumn{1}{l}{Factor} \\
\multicolumn{1}{l}{\MyIndent US data}      & 1.6    & \textbf{1.6} & \textbf{1.6} & 1.9 & 2.1          \\  \hline
\multicolumn{1}{l}{RNN*}\\ 
\multicolumn{1}{l}{\MyIndent Best}      & 1.8   & 2.3 & 2.0 & 2.0 & 1.9\\ 
\multicolumn{1}{l}{\MyIndent Mean Forecast}      & 1.7   & 1.7 & 1.7 & 1.7 & 1.7\\
\multicolumn{1}{l}{\MyIndent Median Forecast}      & 1.7   & 1.7 & 1.7 & 1.7 & 1.7\\\hline
\multicolumn{1}{l}{SPF Median}         & 1.4          & 1.5          & 1.5         & 1.5          & 1.5          \\ \hline\hline
\\ \multicolumn{6}{r}{$^{*}$All RNN models use entire world data cross-section} \\ 
\end{tabular}
%\caption{This table shows the performance of the best, mean, and median forecasts for the performance of gated recurrent units on the test set 2009-Q4 to 2019-Q4. Baseline model performance is also shown.}
\end{table}
% \end{center}

This provides supplementary evidence that the outperformance of our neural network is not due to either over-fitting the test set or over-fitting the architecture choice. Additionally, we ran Monte Carlo simulations (Table \ref{monte_carlo_table_a}) which show that given one hundred random initialization and optimization routines over all five horizons, the model still consistently achieves low root mean squared error and has a low standard deviation -- demonstrating stability and reproducability.

% \begin{center}
% \begin{tabular}[c]{ c  c  c  c  c  c}
% \multicolumn{5}{ c }{RNN Monte Carlo Simulations}\\
%  Time (Q's Ahead) & 1Q & 2Q & 3Q & 4Q & 5Q\\ 
%  Mean RMSE & 1.9 & 2.0 & 2.1 & 2.3 & 2.2 \\
%  Standard Deviation RMSE & .11 & .13 & .15 & .12 & .11 \\
% \end{tabular}
\begin{table}[htbp] \centering
% \centering
\caption{Expansion RNN Monte Carlo Simulations}
\label{monte_carlo_table_a}
\caption*{The mean and standard deviation of the performance of gated recurrent units on the test set 2009-Q4 to 2019-Q4}
\begin{tabular}{cccccc}
% \multicolumn{6}{ c }{RNN Monte Carlo Simulations}      \\
\hline\hline
\multicolumn{1}{r}{Time (Q's Ahead)}        & 1Q   & 2Q   & 3Q   & 4Q   & 5Q   \\ \hline
\multicolumn{1}{l}{Mean RMSE}              & 1.8   & 1.8  & 1.8  & 1.8  & 1.7 \\
\multicolumn{1}{l}{Std Dev RMSE}           & 0.18   & 0.18 & 0.21 & 0.11 & 0.08 \\
\hline\hline
\end{tabular}
% \caption{This table shows the mean and standard deviation of the performance of gated recurrent units on the test set 2009-Q4 to 2019-Q4. Baseline model performance is the same as in table \ref{rmse_table_a}.}
\end{table}
% \end{center}

\newpage

\noindent \textsc{Department of Economics, University of Michigan, Ann Arbor}\newpage 

\newpage 

\end{document}